\pdfoutput=1 
\documentclass[a4paper,11pt]{article}
\PassOptionsToPackage{table}{xcolor}

\usepackage{jheppub} % for details on the use of the package, please
\usepackage[T1]{fontenc} % if needed

\usepackage{tikz}
\usetikzlibrary{shapes,arrows,shadows}
\usepackage{tangocolors}

\usepackage[all]{xy}
\usepackage[percent]{overpic}
\usepackage{slashed}
\usepackage{wrapfig}
\usepackage{tabu}
\usepackage{diagbox}
\usepackage{mathrsfs,amsmath,amssymb,amsthm,amsfonts,tikz,graphicx,accents,hyperref, color}
\usepackage{dsfont,epiolmec, latexsym, stmaryrd, comment}
\usepackage{slashed,ccaption}
\usepackage{mathrsfs, calligra}
\usepackage{leftidx}
\usepackage{import}
\usepackage{multirow}
\usepackage{amsfonts}
\usepackage{pifont}
\usepackage{tabularx}
\usepackage[utf8]{inputenc}
\usetikzlibrary{intersections,calc}
\usepackage{tikz-3dplot}
\usepackage{ifthen}
\usetikzlibrary{arrows}
\usepackage{amsmath}
\usepackage{xcolor}

\usepackage{caption} 
\DeclareMathOperator{\extdm}{d}
\newcommand{\extd}{\extdm \!}

\usepackage{array}
\hypersetup{ linktoc=all,
    colorlinks, linkcolor={palatinateblue},
    citecolor={brightpink}, urlcolor={amaranth}
}

\graphicspath{{Images/}}

\renewcommand{\d}[1]{\ensuremath{\operatorname{d}\!{#1}}}

\def\sideremark#1{\ifvmode\leavevmode\fi\vadjust{\vbox to0pt{\vss% the remark
 \hbox to 0pt{\hskip\hsize\hskip1em%                          will appear only
 \vbox{\hsize2cm\tiny\raggedright\pretolerance10000%          on the side
 \noindent #1\hfill}\hss}\vbox to8pt{\vfil}\vss}}}%
\DeclareSymbolFont{extraup}{U}{zavm}{m}{n}
\DeclareMathSymbol{\varheart}{\mathalpha}{extraup}{86}
\DeclareMathSymbol{\vardiamond}{\mathalpha}{extraup}{87}
\makeatletter
\renewcommand*{\@fnsymbol}[1]{\ensuremath{\ifcase#1\or \clubsuit \or \vardiamond \or \varheart\or
    \spadesuit\or \mathparagraph\or \|\or **\or \dagger\dagger
    \or \ddagger\ddagger \else\@ctrerr\fi}}
\makeatother

\definecolor{rosy}{RGB}{230,235,252}
\definecolor{myframetitle}{RGB}{90,89,170}
\definecolor{myblocktitle}{RGB}{140,185,249}
\definecolor{mytitle}{RGB}{10,80,26}

\definecolor{darkgreen}{RGB}{27,130,45}
\definecolor{darkblue}{rgb}{0,0,0.3}
\definecolor{darkred}{rgb}{0.7,0,0}

\definecolor{light gray}{RGB}{220,220,220}
\definecolor{dark purple}{RGB}{108,0,217}
\definecolor{pink}{RGB}{190,20,100}
\definecolor{orang}{RGB}{193,63,0}
\definecolor{green}{RGB}{11,98,17}
\definecolor{darkpink}{RGB}{153,0,76}
\definecolor{bluegreen}{RGB}{0,102,102}
\definecolor{greenlagan}{RGB}{0,102,0}
\definecolor{redgreen}{RGB}{102,102,0}
\definecolor{Redgreen}{RGB}{153,76,0}
\definecolor{vividviolet}{rgb}{0.62, 0.0, 1.0}
\definecolor{amaranth}{rgb}{0.9, 0.17, 0.31}
\definecolor{palatinateblue}{rgb}{0.15, 0.23, 0.89}
\definecolor{brightpink}{rgb}{1.0, 0.0, 0.5}
\definecolor{cornflowerblue}{rgb}{0.39, 0.58, 0.93}
\definecolor{deepcarminepink}{rgb}{0.94, 0.19, 0.22}
\definecolor{radicalred}{rgb}{1.0, 0.21, 0.37}
\DeclareFontFamily{OT1}{rsfs}{}

\DeclareFontShape{OT1}{rsfs}{m}{n}{ <-7> rsfs5 <7-10> rsfs7 <10->rsfs10}{} 

\DeclareMathAlphabet{\mycal}{OT1}{rsfs}{m}{n}

\newcommand{\di}{\text{d}}

\newcommand{\eps}{\varepsilon}

\newcommand{\eq}[2]{\begin{equation}
#1\label{#2}
\end{equation}}

\newcommand{\be}{\begin{equation}}
\newcommand{\ee}{\end{equation}}
\makeatletter \@addtoreset{equation}{section}

\newcommand{\GH}{\Gamma}

\newcommand{\eom}{\textrm{\tiny{eom}}}

\begin{document}

\preprint{IPM/P-2019/041}
\rightline{TUW--20--02}

\newcommand{\mytitle}{\Huge{T-Witts from the horizon}}

\title{\centerline{\textbf{\mytitle}}}

\author[a]{H.~Adami}
\author[b]{, D.~Grumiller}
\author[a,c]{, S.~Sadeghian}
\author[a,d]{, M.M.~Sheikh-Jabbari}
\author[b]{and C.~Zwikel}
\affiliation{$^a$ School of Physics, Institute for Research in Fundamental
Sciences (IPM),\\ P.O.Box 19395-5531, Tehran, Iran}
\affiliation{$^b$ Institute for Theoretical Physics, TU Wien, Wiedner Hauptstrasse 8--10/136, A-1040 Vienna, Austria}
\affiliation{$^c$ Wigner Research Centre for Physics, Konkoly-Thege Miklos u. 29-33, 1121 Budapest, Hungary}
\affiliation{$^d$ The Abdus Salam ICTP, Strada Costiera 11, Trieste, Italy}

\emailAdd{hamed.adami@ipm.ir, grumil@hep.itp.tuwien.ac.at, jabbari@theory.ipm.ac.ir, saeedeh.sadeghian@wigner.hu, zwikel@hep.itp.tuwien.ac.at}

\abstract{
Expanding around null hypersurfaces, such as generic Kerr black hole horizons, using co-rotating Kruskal--Israel-like coordinates we study the associated surface charges, their symmetries and the corresponding phase space within Einstein gravity. Our surface charges are not integrable in general. Their integrable part generates an algebra including superrotations and a BMS$_3$-type algebra that we dub ``T-Witt algebra''. The non-integrable part accounts for the flux passing through the null hypersurface. We put our results in the context of earlier constructions of near horizon symmetries, soft hair and of the program to semi-classically identify Kerr black hole microstates.
}
\maketitle

\section{Introduction}\label{S.I}

\newcommand{\NHS}{null hypersurface symmetry}

In this paper we study non-extremal Kerr black holes \cite{Kerr:1963ud} from a near horizon perspective within Einstein gravity \cite{Einstein:1915ca}. 
This work is organized as follows:
\begin{itemize}
\item In section \ref{sec:2} we introduce co-rotating Kruskal--Israel like coordinates for Kerr black holes. 
\item In section \ref{sec:3} we provide fall-off conditions suitable to describe the region around the bifurcation surface, discuss the near bifurcation Killing vector algebra and derive the associated surface charges, which turn out to be finite, integrable and conserved. 
\item In section \ref{sec:4} we consider more generic boundary conditions that only assume the existence of a null hypersurface as (fiducial or actual) boundary as well as additional technical assumptions (co-rotation and Taylor-expandability). We derive the near null hypersurface Killing vectors that preserve these fall-off conditions and deduce the associated {\NHS} algebra, which contains a T-Witt subalgebra. 
\item In section \ref{sec:5} we construct the surface charge variations associated with the near horizon symmetry generators of section \ref{sec:4}. They are neither integrable nor conserved, in general. We employ the Barnich--Troessaert modified bracket method to separate the integrable part of the charge and establish a generalized charge conservation equation that relates its time-derivative to the non-integrable part, which is interpreted as flux through the null hypersurface. We also discuss ambiguities of the split into integrable and flux-part and fix them such that the modified bracket algebra closes without central extension, thereby recovering precisely the algebra discussed in section \ref{sec:4}. 
\item In section \ref{sec:onshellphasespace} we examine the on-shell phase space and possible redundancies that can be gauge fixed. We verify for generic and special cases (non-expanding or stationary null hypersurfaces) that there is always a precise match between the independent functions on the boundary phase space and the independent functions in the symmetry generators.
\item In section \ref{sec:7} we impose physically well-motivated conditions on the allowed variations and  recover earlier constructions of  near horizon symmetries as special cases of our analysis.
\item In section \ref{sec:8} we conclude with a summary and give an outlook to current and future research directions.
\end{itemize}
%
%             __
%        _   /  |
%       | \  \/_/
%       \_\| / __              
%          \/_/__\           .--='/~\
%   ____,__/__,_____,______)/   /{~}}}
%   -,-----,--\--,-----,---,\'-' {{~}}
%           __/\_            '--=.\}/
%          /_/ |\\
%               \/
%

%%%%%%%%%%%%%%%%%%%%%%%%%%%%%%%%%%%%%%%%%%%%%%%%%%%%%%%%%%%%%5
\section{Kerr geometry and metric near its bifurcate horizon}\label{sec:2}
%%%%%%%%%%%%%%%%%%%%%%%%%%%%%%%%%%%%%%%%%%%%%%%%%%%%%%%%%%%%%5

The Kerr metric in Boyer--Lindquist coordinates 
\begin{equation}\label{Kerr01}
\begin{split}
\d s^{2}= - \frac{\Delta}{\rho^2}\big(\d t-a\sin^2\theta \d\varphi\big)^{2}+\frac{\rho^{2}}{\Delta} \d r^{2}+\rho^{2} \d\theta^{2}+ \frac{(r^{2}+a^{2})^2\sin^2\theta}{\rho^2}\big(\d\varphi-\frac{a}{r^2+a^2}\d t\big)^2
\end{split}
\end{equation}
with
\begin{equation}
  \Delta = r^{2} - 2 M r +a^{2}\qquad\qquad  \rho^{2}= r^{2} + a^{2} \cos^{2}\theta 
\end{equation}
depends parametrically on mass $M$ and angular momentum $J=M a$. The angular coordinates (also called ``angular part'') are the usual polar angle $\theta\in[0,\pi]$ and the azimuthal angle $\varphi\sim\varphi+2\pi$. The time and radial coordinates (also referred to as ``spacetime part'') are non-compact, $t\in(-\infty,\,\infty)$ and $r\in[0,\,\infty)$. 

The inner and outer horizon radii $r_\pm$ are respectively the smaller and bigger roots of $\Delta=0$,
\be
r_{\pm}=M\pm\sqrt{M^2-a^2}\,.
\ee
The loci $r=r_\pm$ are bifurcate Killing horizons with a bifurcation 2-sphere $\cal B$. The outer one is simultaneously a black hole event horizon and will be our main region of interest.  Surface gravity at the black hole horizon is given by
\begin{equation}
\kappa=\frac{r_+-r_-}{2(r_++r_-)r_+}\,.
\end{equation}
Since Boyer--Lindquist coordinates are not well-adapted to near horizon expansions we introduce new coordinates below.

\subsection{Near horizon coordinates}\label{sec:2.1}

For near horizon analyses various coordinate choices have been used in the literature, e.g.~Eddington--Finkelstein (EF) types of coordinates \cite{Donnay:2015abr, Donnay:2016ejv, Chandrasekaran:2018aop}, conformal types of coordinates (HHPS) \cite{Haco:2018ske} and Rindler coordinates \cite{Grumiller:2019fmp}. Of course, there is also the venerable set of Kruskal coordinates \cite{Kruskal:1959vx} or simplifications thereof \cite{Israel:1966}. Each of these have their own merits and drawbacks. For our purpose the most striking difference between various choices is the way the horizon is approached and whether or not the bifurcation 2-sphere can be covered. We found none of these coordinate systems suitable for our purposes.

Thus, we introduce a Kruskal--Israel-inspired coordinate system that is well-adapted to study not only the black hole horizon, but specifically the region near the bifurcation 2-sphere. The main features of our coordinates are
\begin{itemize}
    \item co-rotation with the horizon (as opposed to Kruskal- or Israel-coordinates)
    \item no mixing of angular and spacetime coordinates (as opposed to HHPS)
    \item cover open region around the bifurcation 2-sphere (as opposed to EF or Rindler)
\end{itemize}
We describe now explicitly the new coordinates, starting with the angular part.

Since we are interested in an expansion around the outer horizon, $r\sim r_+$, it is useful to transform the azimuthal angle so that our coordinate frame is co-rotating with the outer horizon
\eq{
\phi ={\varphi} - \Omega_{\textrm{\tiny H}} \,t\qquad\qquad  \Omega_{\textrm{\tiny H}}= \frac{a}{r_+^2+a^2}\,.
}{eq:kerr2}
As we are content with the polar angle $\theta$ we do not transform it.

We address now the spacetime part. We introduce Kruskal--Israel-like coordinates that we denote by $x^\pm$, defined by
\begin{equation}\label{x-pm-NH-coord}
    x^+=\pm\sqrt{\left|\frac{r-r_{+}}{r_{+}-r_{-}}\right|}\, e^{\kappa t}\qquad\qquad x^-=\pm\sqrt{\left|\frac{r-r_{+}}{r_{+}-r_{-}}\right|}\, e^{-\kappa t}
\end{equation}
the inverse of which is
\begin{equation}\label{t-r-xppm-inverse}
    t= \frac{1}{2 \kappa}\, \ln{\left|\frac{x^+}{x^-} \right|} \qquad\qquad r=r_+ - (r_+ - r_-)\, x^+x^-
\end{equation}
and hence 
\begin{equation}\label{Delta+-}
    \Delta=(r_+ - r_-)^2 x^+x^- (x^+x^- - 1)\qquad\qquad \rho^2=(r_+ - (r_+-r_-) x^+ x^-)^2 + r_+r_-\cos^2\theta\,.
\end{equation}
In this coordinate system $x^\pm \in \mathbb{R}$. The signs in \eqref{x-pm-NH-coord} are fixed as follows. As in the usual Kruskal coordinates, the outside region corresponds to $x^+>0, x^-<0$, the black hole region to $x^\pm>0$, the white hole region to $x^\pm<0$ and the second outside region to $x^+<0, x^->0$. The locus $x^+x^-=0$ describes the bifurcate black hole horizon and $x^+=x^-=0$ the bifurcation 2-sphere $\cal B$. The inner (Cauchy) horizon corresponds to $x^+x^-=1$ and our coordinates break down on it. Thus, the only restriction on the range of the coordinates $x^\pm$ is that their product is smaller than unity, $x^+x^-<1$. 

In these new coordinates the Kerr black hole is described by metrics of the form
\begin{multline}
\extd s^2 = \rho^2\extd\theta^2 - \rho^2\frac{(x^-\extd x^+ + x^+\extd x^-)^2}{x^+x^-(1-x^+x^-)} + \frac{\sin^2\theta}{\rho^2}\Big(  a \big(x^+\extd x^- - x^-\extd x^+\big)\big(2r_+ - (r_+-r_-)x^+x^-\big) \\
+ \big(r_+^2(1-x^+x^-)^2+r_-^2(x^+x^-)^2+r_+r_-(1+2x^+x^-(1-x^+x^-))\big)\,\extd\phi \Big)^2 \\
+ \frac{1-x^+x^-}{\rho^2\,x^+x^-}\, \Big(\rho_+^2\,\big(x^-\extd x^+-x^+\extd x^-\big)-a\sin^2\theta\,(r_+-r_-)\,x^+x^-\extd\phi\Big)^2
    \label{eq:kerr1}
\end{multline}
with $a=\sqrt{r_+r_-}$ and $\rho_+^2=r_+(r_+ + r_-\cos^2\theta)$. In the extremal case, $r_+=r_->0$, the metric \eqref{eq:kerr1} simplifies to the near horizon extremal Kerr (NHEK) geometry \cite{Bardeen:1999px}
\begin{multline}
\extd s^2_\textrm{\tiny{NHEK}} = r_+^2\,(1+\cos^2\theta)\,\Big( - \frac{(x^-\extd x^++x^+\extd x^-)^2}{x^+x^-(1-x^+x^-)} + \frac{(x^-\extd x^+-x^+\extd x^-)^2(1-x^+x^-)}{x^+x^-} \\
+\extd\theta^2 + \frac{4\sin^2\theta}{(1+\cos^2\theta)^2}\,\big(\extd\phi-x^-\extd x^++x^+\extd x^-\big)^2\Big)
\label{eq:EK}
\end{multline}
%%%
%%% Luke, use the source!
%%%
%%% ADDITIONAL INFO ON NHEK
%%%
%%% using the coordinates $\tau=\ln(x^+/x^-)$ and $\rho=x^+x^-$ the NHEK metric no longer is regular at the horizon, but simplifies to
%%%\begin{multline}
%%% \extd s^2 = r_+^2\,(1+\cos^2\theta)\, \Big(\extd\theta^2 - \frac{\extd\rho^2}{\rho(1-\rho) + \extd\tau^2 \rho(1-\rho) \\
%%% + \frac{4\sin^2\theta}{(1+\cos^2\theta)^2}\, \big(\extd\phi -\rho\,\extd\tau\big)^2\Big)}
%%%\label{eq:easteregg}
%%%\end{multline}
%%% NHEK has four Killing vectors, which in the coordinates above are given by $\partial_\phi$, $\partial_\tau$ and
%%%\eq{
%%% e^{\pm\tau/2\sqrt{\rho(1-\rho)}}\Big(\partial_\rho \pm\frac{1}{1-\rho}\partial_\phi\pm \frac{2\rho-1}{\rho(1-\rho)}\partial_\tau\Big)
%%%}{eq:Killing_NEHK}
The reason why we obtain NHEK rather than extremal Kerr is because the coordinate transformation \eqref{t-r-xppm-inverse} is singular in the extremal case and zooms into the region $r=r_+$ for any finite values of $x^\pm$. So, our coordinate system captures generic and near extremal cases. For convenience we collect the ranges of the coordinates: $x^\pm\in\mathbb{R}$, $x^+x^-<1$, $\theta\in[0,\pi]$ and $\phi\sim\phi+2\pi$.

The Killing vectors of the Kerr metric in the Boyer--Lindquist coordinates \eqref{Kerr01} are $\partial_t$ and $\partial_\varphi$, associated with stationarity and axi-symmetry, respectively. The combination $\zeta_{\textrm{\tiny H}}=\partial_t+\Omega_{\textrm{\tiny H}}\partial_\varphi$ is null, tangential and normal at the black hole horizon $r=r_+$, so we use it instead of $\partial_t$. In our new coordinates \eqref{eq:kerr1} the Killing vectors read 
\be\label{Zeta01}
\zeta_{\textrm{\tiny H}}=\kappa\, \big(x^{+}\partial_{+}-x^{-}\partial_{-}\big)\qquad\qquad \zeta_\phi = \partial_{\phi}\,.
\ee 
Consistently, $\zeta_{\textrm{\tiny H}}$ is null at the bifurcate horizon $x^+x^-=0$ and vanishes at the bifurcation 2-sphere $\cal B$, whose properties we discuss next.

\subsection{Bifurcation 2-sphere}\label{se:bi}

The bifurcation surface ${\cal B}$ is a compact surface of $S^2$ topology with metric
\begin{equation}\label{Horizon-metric}
     \d\sigma^2_{_{\cal B}}=  \frac{ (2 M r_{+})^{2} \sin^{2}\theta}{\rho_{+}^{2}} \d\phi^{2}+\rho_{+}^{2} \d\theta^{2}
\end{equation}
where 
\begin{equation}\label{rho-def}
\rho_+^2=r_+(r_++r_-\cos^2\theta)\,.
\end{equation}
The area of the horizon ${\cal B}$ is 
\begin{equation}\label{Horizon-Area}
A=\int_{{\cal B}}\ \d\theta \d\phi\, (2M r_+)\sin\theta=8\pi M r_+\,. \end{equation}

The metric \eqref{Horizon-metric} can be written as conformal to the metric of round 2-sphere
\begin{equation}\label{B-metric-zzbar}
    \d \sigma^{2}_{_{\cal B}} =\Omega_{\text{\tiny{Kerr}}}\, \frac{4 \d z \d{ \bar{z}}}{(1+z \bar{z})^{2}}\,.
\end{equation}
by a Weyl factor 
\eq{
\Omega_{\text{\tiny{Kerr}}}=\frac{( 2Mr_+)^2\sin^2\theta}{4\rho_{+}^{2}} \Big(\mu(\theta)+\frac{1}{\mu(\theta)}\Big)^2
}{eq:kerr3}
with coordinates defined by
\eq{
z= e^{i\phi} \mu (\theta)\qquad\qquad \bar{z}= e^{-i\phi} \mu (\theta)\qquad\qquad   \mu (\theta)= e^{-\frac{r_-}{2M} \cos{\theta}} \cot{\frac{\theta}{2}}\,.
}{zzbar}

%#########################################################################
\section{Near horizon symmetries at the bifurcation surface}\label{sec:3}
%%%%%%%%%%%%%%%%%%%%%%%%%%%%%%%%%%%%%%%%%%%%%%%%%%%%%%%%%%%%%%%%%%%%%%%%%%
In this section we analyse the horizon symmetries and associated charges around the bifurcation surface ${\cal B}$, which we refer to as ``near bifurcation symmetries''. In section \ref{sec:3.1} we discuss the most general fall-off behavior for metric fluctuations preserving the bifurcation surface. In section \ref{sec:3.2} we derive the near bifurcation Killing vectors compatible with the fall-off conditions. In section \ref{se:3.2too} we study the ensuing near bifurcation Killing vector algebra. In section \ref{sec:3.4} we determine the charges associated with the near bifurcation Killing vectors, investigate their integrability and present the algebra generated by them. 

%%%%%%%%%%%%%%%%%%%%%%%%%%%%%%%%%%%%%%%%%%%%%%%%%%%%%%%%%%%%%%%%%%%%%%%%%%
\subsection{Fall-off conditions}\label{sec:3.1}
%%%%%%%%%%%%%%%%%%%%%%%%%%%%%%%%%%%%%%%%%%%%%%%%%%%%%%%%%%%%%%%%%%%%%%%%%%

To motivate our fall-off conditions, we expand the Kerr metric \eqref{eq:kerr1} near the bifurcation surface ${\cal B}$ at $x^\pm=0$,
\begin{multline}\label{ME01}
    \hspace*{-4mm} \d s^2= -4  \rho_{+}^{2} \d x^{+} \d x^{-}
       -{8 M   a \left( \frac{r_{+}^2}{{\rho_{+}^{2}}}+ \kappa r_{+}\right) \sin^{2}\theta } \,\big(x^- \d x^+  -  x^+ \d x^- \big)\d\phi%\\ 
       + \d \sigma^2_{_{\cal B}} +\cdots
\end{multline}
where $\d \sigma^2_{_{\cal B}}$ is given in \eqref{Horizon-metric} and the ellipsis denotes higher order terms in $x^\pm$. In this expansion we are assuming $x^+$ and $x^-$ to be small and of the same order. 

As suggested by the expansion \eqref{ME01} we postulate the near bifurcation fall-off conditions 
\begin{subequations}
\label{FB01}
    \begin{align}
         g_{\pm \pm}& = \mathcal{O}(x^2) & g_{\pm A} &= x^{\mp} C^{\pm}_{A}(x^{B}) + \mathcal{O}(x^{3}) \\ 
         g_{+-}&= \eta(x^{B})+ \mathcal{O}(x^2) &  g_{AB}&= \Omega_{AB}(x^{C})+\mathcal{O}(x^2)
    \end{align}
\end{subequations}
where $x^{A}=(\theta, \phi)$ denote the coordinates on ${\cal B}$. To avoid clutter we use $\mathcal{O}(x^{n})$ as a shorthand for $\mathcal{O}((x^{\pm})^n)$. The near bifurcation expansion functions, $\Omega_{AB}, C_A,\ \eta$, are not constrained by the Einstein field equations to leading order. The fall-off conditions \eqref{FB01} are  the most general ones (subject to Taylor-expandability) preserving the bifurcation 2-sphere at $x^\pm=0$.

%%%%%%%%%%%%%%%%%%%%%%%%%%%%%%%%%%%%%%%%%%%%%%%%%%%%%%%%%%%%%%%%%%%%%%%%%%
\subsection{Near bifurcation Killing vectors}\label{sec:3.2}
%%%%%%%%%%%%%%%%%%%%%%%%%%%%%%%%%%%%%%%%%%%%%%%%%%%%%%%%%%%%%%%%%%%%%%%%%%

The diffeomorphisms that keep the near bifurcation expansion \eqref{FB01} intact are generated by ``near bifurcation Killing vectors'' given by
\begin{equation}\label{NHKV-NB-case}
    \xi^{\pm} = \pm x^{\pm} T^{\pm} (x^A)+\mathcal{O}(x^{3})\qquad\qquad \xi^{A}= Y^{A}(x^B)+ \mathcal{O}(x^2)\,.
\end{equation}
Under a transformation generated by near bifurcation Killing vector fields \eqref{NHKV-NB-case}, the leading order metric functions transform as
\begin{subequations}\label{metric-pert-variation-NB-case}
\begin{align}
    \delta_{\xi} \eta &= Y^{A} \partial_{A} \eta + \left(T^{+}-T^{-} \right) \eta \label{metric-pert-variation-NB-case-eta}\\
    \delta_{\xi} C^{\pm}_{A} &= Y^B \partial_B C^\pm_A  + C^\pm_B \partial_A Y^B + \left(T^{+}-T^{-} \right) C^{\pm}_{A} \mp \eta \partial_{A} T^{\mp} \label{metric-pert-variation-NB-case-CA}\\
    \delta_{\xi} \Omega_{A B} &= Y^C\partial_C \Omega_{A B} + \Omega_{AC}\partial_B Y^C + \Omega_{BC}\partial_A Y^C
\end{align}
\end{subequations}

In the case above $\eta$, $C_A^\pm$ and $\Omega_{AB}$ are tensors (scalar, 1-forms and 2-tensor, respectively) on ${\cal B}$, while in later sections we shall also encounter tensor-densities. The $\xi^\pm$ in \eqref{NHKV-NB-case} generate $x^A$-dependent scalings in $x^\pm$ directions. As required, these transformations map $x^\pm=0$ to itself and hence keep ${\cal B}$ intact. 

\subsection{Near bifurcation Killing vector algebra}\label{se:3.2too}

Under the usual Lie bracket, the  near bifurcation Killing vector fields \eqref{NHKV-NB-case} satisfy the algebra
\begin{equation}\label{NHKV-algebra-NB-case}
    \left[ \xi(T^{+}_{1},T^{-}_{1},Y^{A}_{1}),\,\xi(T^{+}_{2},T^{-}_{2},Y^{A}_{2})\right]=\xi(T^{+}_{12},T^{-}_{12},Y^{A}_{12})
\end{equation}
where
\begin{equation}\label{NHKV-algebra}
    T^{\pm}_{12}= Y^{A}_{1} \partial_{A} T^{\pm}_{2}-Y^{A}_{2} \partial_{A} T^{\pm}_{1}
    \qquad\qquad Y^{A}_{12}= Y^{B}_{1} \partial_{B} Y^{A}_{2}-Y^{B}_{2} \partial_{B} Y^{A}_{1}\,.
\end{equation}
The generators $T^\pm, Y^A$ are functions on ${\cal B}$. As \eqref{NHKV-algebra} shows,  $Y^A$ generate general 2d diffeomrphisms on ${\cal B}$. The $T^\pm$ transform as scalars under 2d diffeomorphisms, while commuting among themselves and with each other. Thus, the $T^\pm$ generate so-called supertranslations.

Analogously to \cite{Barnich:2009se,Barnich:2010eb} for simplicity we impose a further restriction on the metric, namely conformality to the round $S^2$ of the co-dimension two metric $\Omega_{AB}$,
\begin{equation}\label{2d-metric}
\Omega_{AB}\d x^A\,\d x^B=\Omega\,\gamma_{AB}\,\d x^A\,\d x^B=\Omega\,\frac{4\,\d z\,\d {\bar{z}}}{(1+z\,\bar{z})^2}\,.
\end{equation}
The assumption \eqref{2d-metric} allows us to work with a single function $\Omega$ instead of $\Omega_{AB}$. That is, $\eta, \Omega, C^\pm_A$ are now the functions parametrizing our phase space of this section. At the level of diffeos, this amounts to working with a subclass of the $Y^A$ generating Weyl rescalings of ${\cal B}$, i.e., we are restricting to superrotations rather than generic 2d diffeomorphisms. 

In the coordinates $z,\bar z$ defined in \eqref{2d-metric} the generators expand as \begin{subequations}
\label{TTA-expansion}
\begin{align}
&\mathcal T^\pm_{n,m} := \pm z^n \bar z^m x^\pm \partial_\pm&  \xi^\pm\partial_\pm&=\sum_{n,m\in\mathbb{Z}} \tau^{\pm}_{nm}  \mathcal T^\pm_{n,m} \\
&\mathcal L_n := -z^{n+1}\partial_z\ ,\quad {\bar {\mathcal  L}}_n=-\bar z^{n+1} \partial_{\bar z} &
Y^A(x^B)\partial_A &= \sum_{n\in\mathbb{Z}} \big({\cal Y}_n \mathcal L_n+\bar{\cal Y}_n { \bar{\mathcal L}}_n\big) 
\end{align}
\end{subequations}
where $\tau^{\pm}_{nm}$, ${\cal Y}_n$ and $\bar{\cal Y}_n$ are arbitrary numbers. 

In the basis spanned by ${\cal L}_n,{\cal {\bar L}}_n,{\cal T}^\pm_{n,m}$ the algebra \eqref{NHKV-algebra-NB-case} takes the form
\begin{subequations}
\begin{align}\label{NHA-B}
   [{\mathcal L}_n,{\mathcal L}_m ]&= (n-m)\mathcal L_{n+m} & [ {\bar {\mathcal  L}}_n, {\bar {\mathcal  L}}_m ] &= (n-m){\bar{\mathcal L}}_{n+m} \\
   [\mathcal L_{k},\mathcal T^{\pm}_{n,m} ] &=-n\ \mathcal T^{\pm}_{n+k,m} & [ {\bar {\mathcal  L}}_{k},\mathcal T^{\pm}_{n,m} ] &=-m\ \mathcal T^{\pm}_{n,m+k} \\
   [\mathcal L_n,{\bar {\mathcal  L}}_m ]&= 0 & [ \mathcal T^\pm_{n,m}, \mathcal T^\pm_{k,l} ] &=0= [ \mathcal T^+_{n,m}, \mathcal T^-_{k,l} ] \,. 
\end{align}
\end{subequations}
The algebra \eqref{NHA-B} consists of a Witt $\oplus$ Witt algebra (the ``superrotation part''), generated by $\mathcal L_n, {\bar { \mathcal L}}_m$, and two towers of supertranslations generated by $\mathcal T^\pm_{n,m}$. This algebra closely resembles the DGGP algebra \cite{Donnay:2015abr} but we have two copies of supertranslations, instead of one copy there. While the second copy will not play any role in this section, it will become important from section \ref{sec:4} onwards.

%%%%%%%%%%%%%%%%%%%%%%%%%%%%%%%%%%%%%%%%%%%%%%%%%%%%%%%%%%%%%%%%%%%%%%%%%%%%%
\subsection{Conserved charges and their algebra}\label{sec:3.4}
%%%%%%%%%%%%%%%%%%%%%%%%%%%%%%%%%%%%%%%%%%%%%%%%%%%%%%%%%%%%%%%%%%%%%%%%%%%%%

There are different methods to compute the charges and their algebras, including the Hamiltonian formulation \cite{Regge:1974zd, Brown:1986nw, Henneaux:2018cst},  action based  \cite{Lee:1990nz,Iyer:1994ys} or field-equation based \cite{Barnich:2001jy, Barnich:2006av} formulations. These methods provide a formula for computing surface charge variations associated with the near bifurcation Killing vectors, which generically may be written as
\begin{equation}\label{charge-variation-generic}
\slashed{\delta} Q_{\xi}[ g;\, \delta g]=\frac{1}{8\pi G} \int_{\cal B} {\cal Q}^{\mu\nu}[g, \nabla_\mu; \delta g, \xi]\, \d x_{\mu\nu}
\end{equation}
where $ g$ is the metric of the background, $\delta g$ the allowed metric variations around this background and $\xi$ is any near bifurcation symmetry generator. The two-form components ${\cal Q}^{\mu\nu}[g, \nabla_\mu; \delta g, \xi]$ are linear in both $\delta g$ and in $\xi$. Index manipulations and covariant derivative $\nabla_\mu$ are both defined with respect to the background metric $g$. Finally, $\extd\sigma_{\mu\nu}$ denotes the induced co-dimension two volume form on ${\cal B}$. 

The explicit form of ${\cal Q}$ is determined by the theory (and somewhat by the method used as each of the charge-variation computation methods comes with some different ambiguities to be dealt with). The $\slashed{\delta}$ on the left-hand-side of \eqref{charge-variation-generic} is used to stress that $\slashed{\delta} Q_{\xi}[g; \delta g]$ is not necessarily integrable in phase space. That is, there may or may not be a charge $Q_{\xi}[g]$ as a function over the phase space of metrics, such that  $ \delta Q_{\xi}[g]=\slashed{\delta} Q_{\xi}[ g; \delta g]$. If this equality holds we call $\slashed{\delta} Q_{\xi}[ g; \delta g]$ integrable and replace $\slashed{\delta}\to\delta$. 
If the charge is integrable then the fundamental theorem of the covariant phase space method (see \cite{Compere:2018aar} and refs.~therein) states
\begin{equation}\label{charge-bracket-integrable}
  \delta_{\xi_{2}}Q_{\xi_{1}} = \{ Q_{\xi_{1}}, Q_{\xi_{2}} \} = Q_{[\xi_{1},\xi_{2}]}+K(\xi_{1},\xi_{2})
\end{equation}
where the bracket is defined by the first equality and $K(\xi_{1},\xi_{2})$ is a possible central extension.

In the present case the metric $ g + \delta g$ is given by the near bifurcation fall-off \eqref{FB01}, the symmetry generators by the near bifurcation Killing vectors \eqref{NHKV-NB-case} and the theory is general relativity. The action based covariant phase space method \cite{Lee:1990nz,Iyer:1994ys} yields
\begin{equation}\label{CE01}
\slashed{\delta} Q_{\xi}= \frac{1}{8 \pi G} \oint_{{\cal B}} \di^{2}x_{\mu \nu}\sqrt{-g}\, \Big( h^{\lambda [ \mu} \nabla _{\lambda} \xi^{\nu]} - \xi^{\lambda} \nabla^{[\mu} h^{\nu]}_{\lambda} - \frac{1}{2} h \nabla ^{[\mu} \xi^{\nu]} + \xi^{[\mu} \nabla _{\lambda} h^{\nu] \lambda} - \xi^{[\mu} \nabla^{\nu]}h \Big) 
\end{equation}
where $h_{\mu \nu}= \delta g_{\mu \nu}$  is a metric variation allowed by  the near bifurcation fall-off \eqref{FB01}. 

Assuming that the generators $T^\pm$, $Y^A$ are field-independent the charges are integrable and independent from the difference $T^+-T^-$,
\begin{equation}\label{charge-1-NB-case}
    Q(T,\,Y^{A}) \equiv \int_{\cal B}\extd^2x\, \big(T\,{\cal P} + Y^A \, {\cal J}_A\big)
\end{equation}
with $T=(T^+ + T^-)/2$ and the charge densities
\begin{equation}\label{charge-densities-NB-case-1}
{\cal P}= \frac{\Omega}{8 \pi G} \qquad \qquad
{\cal J}_A=-\frac{\Omega}{16 \pi G}\frac{ C^{+}_{A}-C^{-}_{A}}{\eta}\,.
\end{equation}
Since there are fewer charges in \eqref{charge-densities-NB-case-1} than functions parametrizing our phase space, $\eta$, $\Omega$, $C_A^\pm$, there is a redundancy in our phase space. We discuss its physical meaning and how to fix it in section \ref{sec:5.4}. 

As an example we consider Kerr black holes \eqref{eq:kerr1}, the charges of which are given by 
\eq{
{\cal P}^{\textrm{\tiny Kerr}}=\frac{r_+(r_+ + r_-)}{8\pi G}\,\sin\theta \qquad\qquad {\cal J}^{\textrm{\tiny Kerr}}_A = {\cal P}^{\textrm{\tiny Kerr}}\,\eps_A{}^B\partial_B \psi^{\textrm{\tiny Kerr}} 
}{eq:kerrcharges}
with $\eps_\phi{}^\theta = -g^{\theta\theta}\,\sqrt{\det\Omega_{AB}}=-(r_++r_-)\sin\theta/(r_+ + r_-\cos^2\theta)$, and
\eq{
\psi^{\textrm{\tiny Kerr}}=4\arctan\big(\sqrt{r_-/r_+}\cos\theta\big)+2\,\frac{r_+-r_-}{r_++r_-}\,\sqrt{r_-/r_+}\cos\theta
}{eq:psi}
which essentially coincides with Eq.~(VII.1) in \cite{Grumiller:2019fmp}. 

Let us now come back to the general result for the charges \eqref{charge-1-NB-case} and derive the algebra generated by them using \eqref{charge-bracket-integrable}. The transformation laws \eqref{metric-pert-variation-NB-case}  yield
\begin{subequations}
\label{charge-variation-NB-case-P-basis}
\begin{align}
       \delta_{\xi} \mathcal{P}&= Y^{A} \partial_{A} \mathcal{P} +\mathcal{P} \partial_{A} Y^{A} \label{eq:P}  \\
       \delta _{\xi} \mathcal{J}_{A}&= Y^B\partial_B \mathcal{J}_A + \mathcal{J}_B\partial_A Y^B  + \mathcal{J}_A \partial_B Y^B  + \mathcal{P} \partial_{A}T
\end{align}
\end{subequations} 
which is the usual transformation behavior of scalar- and vector-densities of weight one under 2d diffeos generated by $Y^A$. The algebra above coincides with the DGGP algebra \cite{Donnay:2015abr}, see also \cite{Grumiller:2019fmp}, which notably features no central extension, i.e., the quantity $K(\xi_1,\,\xi_2)$ in \eqref{charge-bracket-integrable} vanishes. Thus, the near bifurcation boundary conditions recover known results.

In the next section we generalize these boundary conditions by dropping the assumption of a preserved bifurcation 2-sphere. This is physically well-motivated since real black holes that formed through some collapse of matter do not have bifurcate horizons.

%%%%%%%%%%%%%%%%%%%%%%%%%%%%%%%%%%%%%%%%%%%%%%%%%%%%%%%%%%%%%%%%%%%%%%
\section{T-Witt symmetries on null hypersurfaces}\label{sec:4} 
%%%%%%%%%%%%%%%%%%%%%%%%%%%%%%%%%%%%%%%%%%%%%%%%%%%%%%%%%%%%%%%%%%%%%%

In this section we study near horizon symmetries associated with generic points on the future horizon. For a black hole like Kerr, the future horizon is not just an arbitrary null surface but has more properties and structure, e.g.~it is a Killing and an event horizon. However, these extra properties are not generically respected by physically important processes like objects falling into the black hole or Hawking radiation. To allow for such processes  we relax some of these properties and impose merely the existence of a (co-rotating) null hypersurface with suitable fall-off conditions that we refer to as ``null hypersurface conditions''; the associated symmetries are captured by the {\NHS} algebra.

In section \ref{sec:4.1} we introduce near null hypersurface fall-off behavior, taking into account on-shell constraints. In section \ref{sec:4.2a} we derive the near null hypersurface Killing vectors. In section \ref{sec:4.2b} we introduce the {\NHS} algebra associated with the near null hypersurface Killing vectors, and its particularly important subalgebra, which we call ``T-Witt algebra''. 

%%%%%%%%%%%%%%%%%%%%%%%%%%%%%%%%%%%%%
\subsection{Near null hypersurface expansion and on-shell conditions} \label{sec:4.1}
%%%%%%%%%%%%%%%%%%%%%%%%%%%%%%%%%%%%%%%%%%%%%%%%%%%%%

Consider a null hypersurfaces which with no loss of generality we assume to be located at $x^-=0$. The 1-form normal to the surface, $l_\mu \propto \delta^-_\mu$ at $x^-=0$, by definition has vanishing norm
\begin{equation}\label{g--}
g^{\mu\nu} l_\mu l_{\nu}\big|_{x^-=0} =0  \qquad \Longrightarrow\qquad  g^{--}|_{x^-=0}=0\,.
\end{equation}
The hypersurface is parametrized by $x^+$ and $x^A$, where $x^+$ is the `time' along the null hypersurface and $x^A$ denotes the angular part. As before we assume hypersurface-orthogonality (which in previous sections implied co-rotation) with respect to the null hypersurface
\begin{equation}\label{g-A}
    l^\mu\partial_\mu x^A|_{x^-=0}=0 \qquad \Longrightarrow\qquad g^{-A}\big|_{x^-=0}=0\,.
\end{equation}
A physical consequence of this geometric assumption is that the normal vector field $l^\mu\partial_\mu\propto\partial_+$ has no angular components and hence can be identified with the Hamiltonian (for a suitable choice of the proportionality factor). This justifies to label $x^+$ as `time'. 

The conditions \eqref{g--} and \eqref{g-A} together with regularity of the metric at $x^-=0$ yields the following general fall-off behavior near the null hypersurface
\begin{subequations}
\label{NFH01}
\begin{align}
         g_{+ +} &= {x^- \eta\,{\cal G} }+ \mathcal{O}\left([x^{-}]^{2}\right) & g_{- -} &= \mathcal{F}+  \mathcal{O}\left(x^{-}\right)\\
         g_{+ A} &= x^{-} \mathcal{V}_{A}+\mathcal{O}\left([x^{-}]^{2}\right) & g_{- A} &=\mathcal{U}_{A}+\mathcal{O}\left(x^{-}\right)\\ 
     g_{AB} &= \Omega_{AB}+\mathcal{O}\left(x^{-}\right) & g_{+-}&= \eta+ \mathcal{O}\left(x^{-}\right)
\end{align}
\end{subequations}
where all fields appearing in \eqref{NFH01} are functions of time $x^+$ and angles $x^A$. To have non-vanishing determinant of the metric we require $\eta\neq 0$ and $\det \Omega_{AB} \neq 0$. At this stage it is not clear which (combinations of the) fields correspond to physical boundary degrees of freedom and which of them are pure gauge. We shall clarify this after computing the charges.  As an example the metric expansion of the Kerr black hole metric \eqref{eq:kerr1} around the future null horizon $x^-=0$ is displayed in appendix \ref{Append:Kerr-expansion}.

We consider now on-shell conditions. Unlike the near bifurcation boundary conditions in section \ref{sec:3.1}, the dynamics of the metric functions in \eqref{NFH01} are partly determined by the near null hypersurface Einstein equations, corresponding to the leading orders in $x^-$ of the full field equations $R_{\mu\nu}=0$. The $\mathcal O(1)$ of $R_{++}=0$ in general imposes a relation between time derivatives of  $\det \Omega_{AB}$ and $\det (\partial_+ \Omega_{AB})$. For our choice \eqref{2d-metric},\footnote{%
The physical motivation for the restriction \eqref{2d-metric} is that it eliminates part of the gravitational wave excitations characterized by the traceless part of $\Omega_{AB}$ and thus allows us to focus on the boundary excitations, the main interest of our present work. To be more precise, we are eliminating all outgoing gravitational waves propagating along the null hypersurface and all ingoing gravitational waves that would yield a memory effect, i.e., an imprint on the null hypersurface. The remaining gravitational waves allowed by the restriction  \eqref{2d-metric} are ingoing gravitational waves with a node on the null hypersurface.
} where $\sqrt{\det \Omega_{AB}}=\gamma\Omega$ in which $\gamma$ is a fixed given function of $z, \bar z$, this equation is
\begin{equation}\label{EOM++Omega}
\partial^2_+\Omega+\frac12\partial_+\Omega\,\Big(\GH -\frac{\partial_{+} \eta }{\eta}-\frac{\partial_+\Omega}{\Omega}\Big)=0
\end{equation}
where 
\begin{equation}\label{GH-def}
 \GH:={\cal G}- \frac{\partial_+\eta}{\eta}\,.
\end{equation}
The $\mathcal O(1)$ of $R_{+A}=0$ gives
\begin{equation}\label{EOM-C-A}
  \partial_+\Upsilon_A+{\Omega\partial_A \GH}-\Omega \partial_+\partial_A\ln\Omega+\partial_+\Omega\,\frac{\partial_A\eta}{\eta}=0
\end{equation}
where 
\begin{equation}\label{Upsilon-def}
\Upsilon_A\,[\mathcal V_A,\mathcal U_A,\eta,\Omega]:= -\frac \Omega \eta \,\Big( 
\mathcal V_A -\partial_+ \mathcal U_A+\frac{\partial_+\Omega}{\Omega}\,\mathcal U_A \Big)\,.
\end{equation}
The other components of the field equations do not yield further constraints for the leading order functions.

%%%%%%%%%%%%%%%%%%%%%%%%%%%%%%%%%%%%%%%%%%%%%%%%%%%%%%%%%%%%%%%%%%%%%%%%%
\subsection{Near null hypersurface Killing vectors and field variations}\label{sec:4.2a}
%%%%%%%%%%%%%%%%%%%%%%%%%%%%%%%%%%%%%%%%%%%%%%%%%%%%%%%%%%%%%%%%%%%%%%%%%%
The fall-off conditions \eqref{NFH01}, the null hypersurface at $x^-=0$ and the direction of the 1-form normal $l_\mu$ are preserved by the near null hypersurface Killing vector fields
\begin{subequations}
\label{AKVH01}
\begin{align}
  \xi^{+} &= T^{+}(x^{+},x^{B}){+ x^- \xi_1^+ (x^{+},x^{B})} +\mathcal{O}\left([x^{-}]^{2}\right),\\ \xi^{-}&=- x^{-} T^{-}({x^+},x^{B})+ \mathcal{O}\left([x^{-}]^{2}\right) \\
  \xi^{A} &= Y^{A}(x^{B})+x^-\, \xi_1^A(x^{+},x^{B})+\mathcal{O}\left([x^{-}]^{2}\right)
\end{align}
\end{subequations}
where $Y^A$ satisfy the conformal Killing equation for the metric $\mathcal L_Y \gamma_{AB}=(\bar \nabla_C Y^C ) \gamma_{AB}$ with $\bar{\nabla}_A$ being the covariant derivative with respect to $\gamma_{AB}$, defined in \eqref{2d-metric}. 

The remarkable feature of these symmetry generators is the appearance of two time-dependent functions $T^\pm(x^+, x^A)$.  As discussed below, this time dependence brings novel features and new physics.

Under the transformations generated by the near null hypersurface Killing vector fields \eqref{AKVH01} the functions in the near null hypersurface metric \eqref{NFH01} transform as 
\begin{subequations}
\label{VDF01}
\begin{align}
    \delta_\xi \Omega_{AB} &= T^+ \partial_+  \Omega_{A B} + Y^C\partial_C\Omega_{AB} + \Omega_{AC}\partial_BY^C + \Omega_{BC}\partial_AY^C \\
    \delta_\xi \eta &= \partial_+\big( T^+ \eta \big)- T^- \eta + Y^A \partial_A \eta \label{VDF01-eta}\\
     \delta_\xi \mathcal {G} &= -2\partial_+ T^- +\partial_+(\mathcal{G}\, T^+) + Y^A\partial_A \mathcal G \label{VDF01-G}\,. 
      \end{align}
\end{subequations}
Using \eqref{2d-metric} and that $\gamma_{AB}$ is fixed under the variations, i.e. $\delta_\xi \gamma_{AB}=0$, yields
\begin{equation}\label{delta-Omega}
\delta_\xi\Omega=  T^+\partial_+\Omega+Y^A \partial_A\Omega+\Omega\,\bar{\nabla}_AY^A
\end{equation}
and
\begin{subequations}
\label{VDF02}
\begin{align}
 \delta_\xi \mathcal {F} &= T^+ \partial_{+} \mathcal{F} -2 T^- \mathcal{F} +Y^A \partial_{A} \mathcal{F} +2\, \eta \,\xi^+_1 +2\, \mathcal{U}_{A}\xi^A_1 \\
 \delta_{\xi} \mathcal{U}_{A} &= T^{+} \partial_{+} \mathcal{U}_{A} - T^{-}\mathcal{U}_{A} +\eta \partial_{A}T^{+} +\Omega_{AB}\xi^B_1 
 + Y^B\partial_B \mathcal{U}_A + \mathcal{U}_B \partial_A Y^B \\
 \delta_{\xi} \mathcal{V}_{A} &= \partial_{+} \left( T^{+} \mathcal{V}_{A}\right)  - T^- \mathcal{V}_A -( \eta \partial_A + \mathcal{U}_A\partial_+)T^- + \eta\, \mathcal{G}\,\partial_A T^+ +\Omega_{AB}\partial_+\xi^B_1 \nonumber \\
&\quad +Y^B\partial_B \mathcal{V}_A + \mathcal{V}_B \partial_A Y^B\,.
   \end{align}
\end{subequations}

\newcommand{\G}{\widetilde{\cal G}}

The composite fields which appear in the field equations \eqref{EOM++Omega} and \eqref{EOM-C-A}, $\GH$ and $\Upsilon_A$ defined in \eqref{GH-def} and \eqref{Upsilon-def}
vary as
\begin{align}
    \!\!\delta_\xi \Upsilon_A
    &= T^+\partial_+\Upsilon_A - \Omega\, \GH\,\partial_A T^+  -  \partial_+\Omega\,\partial_AT^+ + 2\Omega\, \partial_A\tilde T^- \nonumber\\
    &\quad + Y^B\partial_B \Upsilon_A + \Upsilon_B\partial_A Y^B + \Upsilon_A \partial_B Y^B 
    \label{Upsilon-variation}\\
  \!\!\delta_\xi  \GH &= T^+\partial_+\GH + \GH\,\partial_+T^+  - 2\partial_+\tilde T^-  +  Y^A\partial_A\GH \label{T-C-G-variation}
\end{align}
with $\tilde T^- := \frac12\,(T^- + \partial_+ T^+)$. The transformation laws above show that $\Upsilon_A$ is a 1-form density of weight one with respect to $Y$-diffeos and mixes with $\GH$ under $T^+$-diffeos. The quantity $\GH$ is a scalar with respect to $Y$-diffeos and a 1-form with respect to $T^+$-diffeos.

\subsection{T-Witt and {\NHS} algebras}\label{sec:4.2b}

As may be anticipated on general grounds (and as shown explicitly in the next section) the charges associated with subleading terms of the near null hypersurface Killing vectors, parameterized by $\xi_1^+, \xi_1^A$ in \eqref{AKVH01}, vanish. Therefore, here we only consider non-zero $T^\pm, Y^A$ in the near null hypersurface Killing vectors \eqref{AKVH01}. Their Lie bracket is
\begin{equation}\label{TT-Lie-bracket}
    \left[ \xi(T^{+}_{1},T^{-}_{1},Y^{A}_{1}),\xi(T^{+}_{2},T^{-}_{2},Y^{A}_{2})\right]=\xi(T^{+}_{12},T^{-}_{12},Y^{A}_{12})
\end{equation}
where
\begin{subequations}\label{ABCD}
\begin{align}
    T^{+}_{12} & =T_{1}^{+} \partial_{+}T_{2}^{+}-T_{2}^{+} \partial_{+}T_{1}^{+}+ Y^{A}_{1} \partial_{A} T^{+}_{2}-Y^{A}_{2} \partial_{A} T^{+}_{1}\\
    T^{-}_{12} & =T_{1}^{+} \partial_{+}T_{2}^{-}-T_{2}^{+} \partial_{+}T_{1}^{-}+   Y^{A}_{1} \partial_{A} T^{-}_{2}-Y^{A}_{2} \partial_{A} T^{-}_{1} \\ 
    Y^{A}_{12}& = Y^{B}_{1} \partial_{B} Y^{A}_{2}-Y^{B}_{2} \partial_{B} Y^{A}_{1}\,.
\end{align}
\end{subequations}

With the Taylor--Maclaurin expansions in $x^+$ and the Laurent expansions in $z$ and $\bar z$
\begin{subequations}\label{BMS-Witt-Witt}
\begin{align}
T^+(x^+; z, \bar z)\,\partial_+ &= \sum_{n=-1}^\infty \sum_{p,q=-\infty}^\infty {\tau}^{n}_{pq}\, {\cal R}_n^{p,q} & {\cal R}_{n}^{p,q} &:= -(x^+)^{n+1} z^p \bar z^q\, \partial_+\\ 
T^-(x^+; z, \bar z)\,\partial_- &= \sum_{n=0}^\infty \sum_{p,q=-\infty}^\infty {\rho}^n_{pq}\, {\cal T}_n^{p,q}  & {\cal T}_n^{p,q} &:= -(x^+)^{n}z^p \bar z^q\, \partial_-\\ 
Y^A(z, \bar z)\,\partial_A &= \sum_{p=-\infty}^\infty \left({\cal Y}_p\, {\cal L}^p + {\bar{\cal Y}}_p\, {\cal {\bar L}}^p \right) & {\cal L}^p &:= -z^{p+1} \,\partial_z\,, \quad {\cal {\bar L}}^p:= -{\bar z}^{p+1}\,\partial_{\bar z}
\end{align}
\end{subequations}
where ${\tau}^n_{pq}, {\rho}^n_{pq}, {\cal Y}_{n}$ and ${\bar{\cal Y}}_{n}$ are some arbitrary numbers, the algebra \eqref{TT-Lie-bracket} reads
\begin{subequations}
\label{eq:angelinajolie}
\begin{align}
   [ {\cal R}_n^{p,q}, {\cal R}_m^{r,s}] &= (n-m)\, {\cal R}_{n+m}^{p+r,q+s}  &
   [ {\cal R}_n^{p,q}, {\cal T}_m^{r,s}] &= -m\, {\cal T}_{n+m}^{p+r,q+s} &
   [ {\cal T}_n^{p,q}, {\cal T}_m^{r,s}] &=0 
\label{Future-horizon-algebra-of-generators-BMS3} \\
   [{\cal L}^p, {\cal L}^q] &=(p-q)\, {\cal L}^{p+q} & [{\bar {\cal L}}^p,{\cal {\bar L}}^q] &=(p-q)\, {\cal {\bar L}}^{p+q} & [{{\cal L}}^p,{\cal {\bar L}}^q]&=0 \label{Future-horizon-algebra-of-generators-Witt-Witt} \\
 [  {\cal L}^r,{\cal R}_{m}^{p,q}] &= -p\, {\cal R}_m^{p+r,q} & [  {\cal {\bar L}}^r, {\cal R}_m^{p,q}] &= -q\, {\cal R}_{m}^{p,q+r}\\
 [  {\cal L}^r,{\cal T}_{m}^{p,q}] &= -p\, {\cal T}_m^{p+r,q} &  [  {\cal {\bar L}}^r, {\cal T}_m^{p,q}] &= -q\, {\cal T}_m^{p,q+r}
 \end{align}
 \end{subequations}
 The Taylor--Maclaurin indices are such that they guarantee smoothness in the $x^+\to 0$ limit, that is $n\geq -1$ for ${\cal R}_n^{p,q}$ and $n\geq 0$ for ${\cal T}_n^{p,q}$.\footnote{We address generalizations of this smoothness condition in the concluding section \ref{sec:8}.} The Laurent indices $p,q,r,s$  are integers.
 
 We refer to the full algebra \eqref{eq:angelinajolie} as ``{\NHS} algebra'' (NHS); it has several interesting subalgebras (see appendix \ref{Appen:B}). We call the one generated by ${\cal R}_n^{p,q}$ and ${\cal T}_n^{p,q}$ ``T-Witt algebra'' \eqref{Future-horizon-algebra-of-generators-BMS3}, and the one generated only by ${\cal R}_n^{p,q}$ ``S-Witt algebra''. At a given point on the bifurcation sphere, the S-Witt algebra reduces to a Witt algebra and the T-Witt algebra reduces to an algebra closely related to the BMS$_3$ algebra \cite{Ashtekar:1996cd,Barnich:2006av}. It differs from the BMS$_3$ algebra only in the bracket $[{\cal R}_n, {\cal T}_m]$ that is proportional to $-m$ instead of $(n-m)$. This algebra was called BMS$_3^{(0)}$ in \cite{Grumiller:2019fmp} and $W(0,0)$ in \cite{Parsa:2018kys}. The subalgebra \eqref{Future-horizon-algebra-of-generators-Witt-Witt}  generated by ${\cal L}^p$ and ${\cal L}^q$ yields superrotations in the sense of \cite{Barnich:2009se, Barnich:2011ct}. We discuss the relation to previous near horizon and {\NHS} algebras in section \ref{sec:7}, but point out already here that the T-Witt algebra was missed in all previous constructions.

%$$$$$$$$$$$$$$$$$$$$$$$$$$$$$$$$$$$$$$$$$$$$$$$$$$$$$$$$$$$$$$$$$$$$$$$$$$$$$$$$$$$$$
\section{Charges, fluxes and generalized charge conservation} \label{sec:5}
%$$$$$$$$$$$$$$$$$$$$$$$$$$$$$$$$$$$$$$$$$$$$$$$$$$$$$$$$$$$$$$$$$$$$$$$$$$$$$$$$$$$$$$

We first compute the variation of the charges associated to the symmetries \eqref{AKVH01} in section \ref{sec:5.1}. We find out that they are not integrable. One can separate the integrable and non-integrable parts of the charge variation. There is, however, an ambiguity in doing so. In section \ref{sec:5.2} we use the Barnich--Troessaert modified bracket method \cite{Barnich:2011mi} to fix this ambiguity by demanding the absence of a central extension in the {\NHS} algebra. Section \ref{sec:5.3} establishes a generalized charge conservation equation relating the non-integrable part to the time-derivative of the integrable part of the charge.

\subsection{Charge variations and non-integrability of the charge}\label{sec:5.1}

Inserting the near null hypersurface expansion \eqref{NFH01} together with the near null hypersurface Killing vectors \eqref{AKVH01} into the surface charge variations \eqref{CE01} yields
\begin{multline}\label{Non-I-Ch}
  \slashed{\delta} Q_{\xi} =  \int_{\gamma} \,  \delta \big(\Omega \partial_{+}\ T^{+} - 2T^{+} \partial_+ \Omega + \Omega\, T^{-}
    +Y^{A} \Upsilon_A\big)   \\ 
    + \int_{\gamma} \, T^{+}\left[ 
 \left(-\GH+\frac{\partial_+ \Omega}{\Omega}\right)\delta \Omega +\frac{\partial_+ \Omega}{\eta} \delta \eta \right]
\end{multline}
where we used the conformal form of the angular part, so that the integrals are over a 2d space with reference metric $\gamma_{AB}$ defined in \eqref{2d-metric}. For convenience we absorb an overall factor containing Newton's constant into the definition of the integral.
\begin{equation}\label{INT-DEF}
     \int_{\gamma}\ \bullet\, \equiv  \,\frac{1}{16 \pi G} \int\ \d{}^2x\sqrt{\gamma}\ \bullet
\end{equation}
For later purposes we highlight that the charge variation \eqref{Non-I-Ch} is independent from $\xi_1^+$ and from $\xi_1^A$. 

While the first line in \eqref{Non-I-Ch} is manifestly  the variation of a function over the phase space, the second line is not. 
Therefore, we split \eqref{Non-I-Ch} 
\begin{equation}
\slashed{\delta} Q_{\xi} = \int_\gamma \left( \delta \mathcal  Q^{\text{I}}_{\xi} +\mathcal F_\xi(\delta g) \right)
\end{equation}
into integrable part
\begin{equation}\label{Q-I-int}
           \mathcal{Q}^{\text{I}}_{\xi}=\Omega \partial_{+} T^{+} - 2T^{+} \partial_+ \Omega + \Omega\, T^{-}  +Y^{A}\Upsilon_A
\end{equation}
and non-integrable part
\begin{equation}\label{cal-F}
    \mathcal{F}_{\xi} (\delta g) =T^{+}  \, \bigg[\bigg(-\GH+\frac{\partial_+ \Omega}{\Omega}\bigg)\delta \Omega +\frac{\partial_+ \Omega}{\eta} \delta \eta
 \bigg]\,.
\end{equation}
This split is not unique due to the ambiguity  
\begin{equation}\label{Amb-01}
\mathcal{Q^{\text{I}}}_{\xi}\to  \mathcal{ \tilde{Q}}^{\text{I}}_{\xi}= \mathcal{Q^{\text{I}}}_{\xi}+\mathcal A_{\xi}(g)\qquad \qquad {\mathcal F}_{\xi} (\delta g)\to {\cal {\tilde{F}}}_{\xi} (\delta g)={\mathcal F}_{\xi} (\delta g)- \delta \mathcal A_{\xi} (g)
\end{equation}
for any arbitrary functional $\mathcal A_{\xi}(g)$. We shall discuss below how this ambiguity can be fixed.

\subsection{Barnich--Troessaert method and modified bracket algebra} \label{sec:5.2}

As discussed in the first two paragraphs of section \ref{sec:3.4}, for the integrable charges the fundamental theorem of the covariant phase space method implies that the algebra of charges is the same as the algebra of asymptotic (near bifurcation surface/near horizon/near null hypersurface) Killing vectors (up to a possible central extension). One may use a similar criterion for fixing the ambiguity in separation of the integrable and non-integrable parts when charges are not integrable, as in the case we are dealing with. This method was developed by Barnich and Troessaert in  \cite{Barnich:2011mi}. They proposed a ``modified bracket'' of the integrable part of charges, $Q_{\xi}^{\text{\tiny{BT}}}$, such that\footnote{One should note the difference between ``adjusted brackets'' which arise due to field dependence of the generators, e.g. see \cite{Compere:2015knw}, and the ``modified bracket'' \cite{Barnich:2011mi} which arise in the non-integrability context.}
\begin{equation}\label{MB-1}
\{Q_{\xi_{1}}^{\text{\tiny{BT}}},Q_{\xi_{2}}^{\text{\tiny{BT}}}\}^*:=\delta_{\xi_{2}} Q^{\text{\tiny{BT}}}_{\xi_{1}} + F^{\text{\tiny{BT}}}_{\xi_{2}} (\delta_{\xi_{1}} g) 
\end{equation}
where $Q_{\xi_{1}}^{\text{\tiny{BT}}}=\int_\gamma \mathcal Q_\xi^{\text{\tiny{BT}}}$ and $F^{\text{\tiny{BT}}}_\xi=\int_\gamma \mathcal F^{\text{\tiny{BT}}}_\xi $. On-shell the modified bracket satisfies 
\begin{equation}\label{modifiedbracket}
\{Q_{\xi_{1}}^{\text{\tiny{BT}}},Q_{\xi_{2}}^{\text{\tiny{BT}}}\}^*= Q^{\text{\tiny{BT}}}_{[\xi_{1},\xi_{2}]} +K_{\xi_{1},\xi_{2}} 
\end{equation}
where $K_{\xi_{1},\xi_{2}}$ is a central extension term. In our case on-shell the latter is given by
\begin{equation}\label{CET-01}
K_{\xi_{1},\xi_{2}} = -\int_\gamma \Omega  \left[\left( T_{1}^{+} \partial_{+}^{2} T_{2}^{+}-T_{2}^{+} \partial_{+}^{2} T_{1}^{+} \right)
   +\left( T_{1}^{+} \partial_{+} T_{2}^{-}-T_{2}^{+} \partial_{+} T_{1}^{-} \right)\right] 
\end{equation}
The central extension \eqref{CET-01} is in general non-vanishing.

Under the ambiguity \eqref{Amb-01} the form of the modified bracket \eqref{modifiedbracket} is preserved but the central extension term shifts,
\begin{equation}\label{CET-02}
   \tilde{K}_{\xi_{1},\xi_{2}}= K_{\xi_{1},\xi_{2}} + \delta_{\xi_{2}} A_{\xi_{1}}(g)- \delta_{\xi_{1}} A_{\xi_{2}}(g)- A_{[\xi_{1},\xi_{2}]}(g)
\end{equation}
where $A_\xi(g)=\int_\gamma \mathcal A_\xi(g)$. This begs the question whether or not there is a choice for the ambiguity such that the central extension vanishes. We show now that the answer is yes. To be minimalistic we consider only ambiguities that are linear combinations of terms resembling the ones already present in the non-integrable part \eqref{cal-F}, namely $T^+(\partial_+ \Omega) \log\Omega$, $T^+(\partial_+ \Omega) \log \eta$ and $T^+\,\GH$.  We find that the choice
\begin{equation}\label{Amb-MB}
\mathcal A_\xi(g) =-T^+\,\Omega\,\GH
\end{equation}
for the ambiguity leads to vanishing central extension, $\tilde K_{\xi_{1},\xi_{2}}=0$. (The choice \eqref{Amb-MB} is not unique; for instance, adding to $\mathcal A_\xi(g)$ a term proportional to $T^+\partial_+\Omega$ would not change our conclusions; however, the choice above is minimalistic.) 

Thus, we fix the ambiguity by demanding the absence of a central extension. Using \eqref{Amb-MB} the integrable and non-integrable parts of the charge become
\begin{align} \label{QI-MB}
 Q^{\text{\tiny{BT}}}_{\xi} &=  \int_\gamma \Big[\big(\partial_{+} T^{+}+T^{-}  \big)\ \Omega- T^{+} \big(2\partial_+ \Omega + \Omega \,\GH \big)  +Y^{A}\Upsilon_A\Big] \\
\label{F-MB}
F^{\text{\tiny{BT}}}_{\xi} (\delta g) &= \int_\gamma
T^+ \, \bigg[ 
\frac{\partial_+ \Omega}{\Omega}\delta \Omega +\frac{\partial_+ \Omega}{\eta} \delta \eta +\Omega \delta\GH
 \bigg]\,.
\end{align}
Introducing modes for the charges analogous to \eqref{BMS-Witt-Witt},
\begin{subequations}
    \label{eq:Qmodes}
\begin{align}
    R_n^{p,q} &:=  Q^{\text{\tiny{BT}}}_{\xi} && \textrm{for}\qquad  \xi = - (x^+)^{n+1}z^p\bar z^q \partial_+  \\
    T_n^{p,q} &:= Q^{\text{\tiny{BT}}}_{\xi} && \textrm{for}\qquad  \xi = - x^-(x^+)^nz^p\bar z^q \partial_- \\
    L^p &:= Q^{\text{\tiny{BT}}}_{\xi} &&\textrm{for}\qquad  \xi = -z^{p+1}\partial_z \\
    \bar L^p &:=Q^{\text{\tiny{BT}}}_{\xi} && \textrm{for}\qquad  \xi = -\bar z^{p+1}\partial_{\bar z} 
\end{align}
\end{subequations}
their modified bracket algebra
\begin{subequations}
    \label{eq:keyresult}
\begin{align}
   \{R_n^{p,q},\, R_m^{r,s}\}^\ast &= (n-m)\, R_{n+m}^{p+r,\,q+s}  &
   \{R_n^{p,q},\, T_m^{r,s}\}^\ast &= -m\, T_{n+m}^{p+r,\,q+s} \\
   \{T_n^{p,q}, T_m^{r,s}\}^\ast &=0 &   \{L^p,\,{\bar L}^q\}^\ast&=0 \\
   \{L^p,\, L^q\}^\ast &=(p-q)\, L^{p+q} & \{{\bar L}^p,\,{\bar L}^q\}^\ast &=(p-q)\,  {\bar L}^{p+q}  \\
 \{L^r,\, R_m^{p,q}\}^\ast &= -p\, R_m^{p+r,\,q} & \{{\bar L}^r,\, R_m^{p,q}\}^\ast &= -q\, R_m^{p,q+r}\\
 \{L^r,\, T_m^{p,q}\}^\ast &= -p\, T_m^{p+r,\,q} &  \{{\bar L}^r,\, T_m^{p,q}\}^\ast &= -q\, T_m^{p,\,q+r}
\end{align}
\end{subequations}
coincides with the {\NHS} algebra \eqref{eq:angelinajolie} without central extension. This is one of the key results of our paper.

\subsection{Generalized charge conservation, Hamiltonian and entropy}\label{sec:5.3}

We study now the time evolution of the charge $Q^{\text{\tiny{BT}}}_\xi$ \eqref{QI-MB}. Depending on the observer, there are different interesting cases for the choice of time coordinate. Below, we discuss two such cases.

\paragraph{Null hypersurface observer ($\boldsymbol{x^+}$ as time).}
The natural time coordinate for an observer temporarily residing at the $x^-=0$ null hypersurface is $x^+$. The corresponding symmetry generator is translation along $x^+$, ${\cal R}^{0,0}_{-1}=-\partial_+$ and the corresponding charge \eqref{QI-MB}, the Hamiltonian, is $Q^{\text{\tiny{BT}}}_{\partial_+}=R^{0,0}_{-1}=\int_\gamma(2\partial_+\Omega+\Omega\GH)$. The charge algebra \eqref{eq:keyresult} implies that this Hamiltonian commutes with superrotations $L^r, {\bar L}^s$ but does not commute with the T-Witt generators $R_n^{p,q}, T_n^{p,q}$. In the stationary limit, where $\partial_+\Omega=\GH=0$, the Hamiltonian $Q^{\text{\tiny{BT}}}_{\partial_+}$ vanishes and is thus not a useful observable. So for (almost) stationary situations one should use the time coordinate below instead. 

Nevertheless, if one considers the charge associated with the Hamiltonian $Q^{\text{\tiny{BT}}}_{\partial_+}$ and evaluates its bracket with another charge $Q^{\text{\tiny{BT}}}_\xi$ for generic $\xi$ then \eqref{modifiedbracket} yields
\begin{equation}
\{Q^{\text{\tiny{BT}}}_{\partial_+},{Q}_{\xi}^{\text{\tiny{BT}}}\}^*={ Q}_{[\partial_+,\xi]}^{\text{\tiny{BT}}}={Q}_{\partial_+\xi}^{\text{\tiny{BT}}}
\end{equation}
while \eqref{MB-1} yields
\begin{equation}
\{ Q_{\partial_+}^{\text{\tiny{BT}}},{Q}_{\xi}^{\text{\tiny{BT}}}\}^*=-\delta_{\partial_+} Q_\xi^{\text{\tiny{BT}}}-{ F}^{\text{\tiny{BT}}}_{\partial_+}(\delta_\xi g)\,.
\end{equation}
As implied by \eqref{VDF01} and \eqref{delta-Omega}, we have $Q_{\partial_+\xi}^{\text{\tiny{BT}}}+\delta_{\partial_+} Q_\xi^{\text{\tiny{BT}}}={\partial_+} Q_\xi^{\text{\tiny{BT}}}$ and therefore arrive at the generalized charge conservation equation (GCCE)
\begin{equation}
{\partial_+} Q_\xi^{\text{\tiny{BT}}}=-{ F}^{\text{\tiny{BT}}}_{\partial_+}(\delta_\xi g)\,. 
\label{GCCE}
\end{equation}
The GCCE means that the non-integrable part of the charge is the source of the non-conservation of the integrable part and therefore represents the flux passing through the null hypersurface. The GCCE \eqref{GCCE} provides a consistency check of our boundary conditions and is expected to hold on general grounds \cite{Barnich:2011mi}, see also \cite{Bunster:2018yjr} for the canonical viewpoint and \cite{Donnay:2016ejv} for similar discussions and results in a special case addressed in section \ref{sec:7.2}. 

\paragraph{Stretched horizon observer ($\boldsymbol{t=\frac{1}{2\kappa}\ln(-x^+/x^-)}$ as time).}  For an observer just outside the horizon in the stationary case the most natural time-translation generator is the one associated with the Killing vector $\zeta_{\textrm{\tiny{H}}}$ given in \eqref{Zeta01}. The corresponding time coordinate is proportional to $\ln(-x^+/x^-)$, see \eqref{t-r-xppm-inverse}. Even for non-stationary situations we can consider this time-coordinate. Following \cite{Grumiller:2019fmp} we define the near horizon Hamiltonian $H_{\text{\tiny NH}}$ as 
\eq{
H_{\text{\tiny NH}} := Q_{\partial_t}^{\text{\tiny{BT}}} = \kappa\, \big(T_0^{0,0} - R_0^{0,0}\big)\,. 
}{eq:H}
The Hamiltonian $H_{\text{\tiny NH}}$ commutes with superrotations and all $T^{p,q}_n$. In the stationary case it is conserved and commutes with all other charges, and hence the associated physical excitations are soft hair, as explained in \cite{Grumiller:2019fmp}.  For the Kerr background (see section \ref{sec:2}), the near horizon Hamiltonian is given by the Wald entropy times Hawking temperature, thereby recovering the near horizon first law \cite{Donnay:2015abr}.

In full generality, however, $H_{\text{\tiny NH}}$ is not conserved since $H_{\text{\tiny NH}}$ does not commute with $R^{p,q}_n$. This non-conservation in particular, and time variation of the charges in general, is captured by a GCCE similar to \eqref{GCCE}:
\begin{equation}\label{entropy-flux}
{\partial_t} Q_\xi^{\text{\tiny{BT}}}=-{ F}^{\text{\tiny{BT}}}_{\partial_t}(\delta_\xi g)\,. 
\end{equation}
In the adiabatic limit, the left hand side of the above equation for $\xi=\zeta_{\textrm{\tiny{H}}}$ \eqref{Zeta01} is the time variation of the entropy, which equals to the flux through the null hypersurface. We shall return to this point in section \ref{sec:8}.

%$$$$$$$$$$$$$$$$$$$$$$$$$$$$$$$$$$$$$$$$$$$$$$$$$$$$$$$$$$$$$$$$$$$$$$$$$$$$$$$$$$$$$
\section{Redundancies and on-shell phase space}\label{sec:onshellphasespace} 
%$$$$$$$$$$$$$$$$$$$$$$$$$$$$$$$$$$$$$$$$$$$$$$$$$$$$$$$$$$$$$$$$$$$$$$$$$$$$$$$$$$$$$
 
The phase spaces defined in the previous three sections describe boundary excitations [since we switched off the bulk excitations through our assumption \eqref{2d-metric}]. These boundary excitations correspond to edge states and are physical, modulo residual pure gauge redundancies, so that the number of functions parametrizing these phase spaces are typically larger than the number of surface charges. We have seen this effect already in section \ref{sec:3}: the near bifurcation Killing vectors involve $T^\pm$, while the charges \eqref{charge-1-NB-case} depend only on their sum. So there is a pure gauge degeneracy in the near bifurcation Killing vectors. This particular degeneracy is lifted in sections \ref{sec:4}-\ref{sec:5}. However, even in the phase space defined through the boundary conditions in section \ref{sec:4} and the charge analysis in section \ref{sec:5} there are still gauge redundancies, which we address in this section. 

In section \ref{sec:5.4} we perform residual gauge fixing to remove gauge redundancies from the boundary phase space. In section \ref{sec:6.2} we discuss the on-shell phase space to verify that the number of charges matches with the number of functions parametrizing our boundary phase space. Two special cases require separate discussion, namely non-expanding and sationary null hypersurfaces, which we deal with in section \ref{sec:6.3}.

\subsection{Fixing residual gauge redundancies}\label{sec:5.4}

The first type of pure gauge redundancy arises when certain symmetry generators do not contribute to the charges at all. In the present case the functions $\xi_1^+$ and  $\xi^A_1$ appearing in the near null hypersurface Killing vectors \eqref{AKVH01} do not contribute to the charge variation \eqref{Non-I-Ch}. Therefore, the diffeomorphisms generated by near null hypersurface Killing vectors that have only these functions switched on are proper diffeomorphisms, even at the null hypersurface boundary. Anticipating this result we considered only the remaining free functions, $T^\pm$ and $Y^A$, when deriving the near null hypersurface Killing vector algebra \eqref{eq:angelinajolie}.

There is a second type of pure gauge redundancies: it could be that not all linear combinations of free functions appearing in the near null hypersurface expansion of the metric appear in the charges. Linear combinations that do not appear thus do not label any physical states and can be gauge fixed to a convenient form. In the present case there are four scalar fields, $\eta, {\cal F}, {\cal G}$ and $\Omega$, and two vector fields, ${\cal V}_A$ and ${\cal U}_A$, appearing in our parametrization of the near null hypersurface metric \eqref{NFH01}. However, the expressions for the charges \eqref{QI-MB} and fluxes \eqref{F-MB} depend only on three of the scalar fields --- there is no dependence on $\cal F$ --- and on one combination of the vector fields, namely $\Upsilon_A$ defined in \eqref{Upsilon-def}. This means that ${\cal F}$ and one combination of ${\cal V}_A$ and ${\cal U}_A$ that is linearly independent from $\Upsilon_A$ can be gauge-fixed conveniently.

Recalling \eqref{VDF02}, we eliminate now $\cal F$ and ${\cal U}_A$ by a diffeomorphism generated by a near null hypersurface Killing vector \eqref{AKVH01} with non-zero functions\footnote{To eliminate ${\cal F}$ and ${\cal U}_A$ we need to consider the finite form of diffeomorphisms generated infinitesimally by $\xi_1$. This means in particular that we need to keep the terms to second order in $\xi_1$ in our transformations, whereas \eqref{VDF02} are only showing the first order ones. The transformation for ${\cal F}$ is then modified to $\delta_\xi\mathcal{F} +\Omega_{AB} \xi_1^A\xi_1^B,$ where $\delta_\xi {\cal F}$ is given in \eqref{VDF02}. The expression for $\delta_\xi {\cal U}_A$ does not receive second order corrections.}
\eq{
{\xi_1^+}=-\frac{1}{2\eta}({\cal F}-\Omega^{AB}{\cal U}_A {\cal U}_B)\qquad\qquad \xi_1^A=-\Omega^{AB}{\cal U}_B\,.
}{eq:whatever}

Removing the gauge freedom from our phase space reduces the near null hypersurface expansion \eqref{NFH01} of the metric
\begin{subequations}
\label{NFH-gauge-fixed}
    \begin{align}
         g_{+ +} &= x^- \eta\,{\cal G}+ \mathcal{O}\left([x^{-}]^{2}\right) & g_{- -} &=   \mathcal{O}\left(x^{-}\right)\\
         g_{+ A} &= x^{-} \theta_{A}+\mathcal{O}\left([x^{-}]^{2}\right) & g_{- A} &=\mathcal{O}\left(x^{-}\right)\\     
         g_{AB} &= \Omega_{AB} %+\lambda_{AB} x^-+ \mathcal{O}\left([x^{-}]^{2}\right) 
         + \mathcal{O}\left(x^{-}\right) 
         &  g_{+-}&= \eta+ \mathcal{O}\left(x^{-}\right)
        \end{align}
\end{subequations}
where 
$\theta_A:=%{\cal V}_A-\partial_+{\cal U}_A+\frac{\partial_+\Omega}{\Omega} {\cal U}_A=
-\frac{\eta}{\Omega} \Upsilon_A$. 
The dynamical fields ${\cal G}$, $\eta$, $\Omega_{AB}=\gamma_{AB}\,\Omega$ [see \eqref{2d-metric}] and ${\theta}_A$ are  functions of time $x^{+}$ and angles $x^{A}$ and parameterize our phase space. Terms not displayed are subleading and do not contribute to the charges or fluxes.

Higher powers of $x^-$ in the near null hypersurface Killing vectors \eqref{AKVH01} do not contribute to the charges and thus generate gauge redundancies that can be used to remove the higher powers in the $g_{-\mu}$ components of the metric. The fully gauge-fixed version of the metric \eqref{NFH-gauge-fixed}
\begin{subequations}
\label{metric-gauge-fixed-sec:4}
    \begin{align}
         g_{+ +} &= x^- \eta\,{\cal G}+ \mathcal{O}\left([x^{-}]^{2}\right) & g_{- -} &=  0 \\
         g_{+ A} &= x^{-} \theta_{A}+\mathcal{O}\left([x^{-}]^{2}\right) & g_{- A} &= 0 \\     
         g_{AB} &= \gamma_{AB}\,\Omega + \lambda_{AB}\, x^-+ \mathcal{O}\left([x^{-}]^{2}\right) 
         &  g_{+-}&= \eta\,.
        \end{align}
\end{subequations}
is preserved by near null hypersurface Killing vectors
\eq{
\xi=T^+(x^+,x^B)\,\partial_+ - x^- T^-(x^+,x^B)\,\partial_- + Y^A(x^B)\,\partial_A + \xi^{\textrm{\tiny fix}}\,.
}{eq:nhkv}
The subleading term $\xi^{\textrm{\tiny fix}} = x^-\hat\xi^A\,\partial_A + x^{-\,2}\hat\xi^-\,\partial_- \hat\xi^A$ is completely determined by the metric and the leading order functions, $\hat\xi^A=-\frac{\eta}{\Omega} \gamma^{AB}\partial_B T^+ + {\cal O}(x^-)$ and $\hat\xi^- = -\frac{1}{2\eta}\,\hat\xi^A\big(\theta_A + \partial_A\eta\big) + {\cal O}(x^-)$. This shows explicitly that all residual gauge redundancies present in the near null hypersurface Killing vectors \eqref{AKVH01} are removed and the remaining free functions, $T^\pm$ and $Y^A$, generate null hypersurface symmetries that are not gauge redundancies in general.

After removing all residual gauge redundancies our boundary phase space is therefore characterized by three scalar functions, $\eta, \GH, \Omega$ and a vector $\Upsilon_A$; the quantity $\lambda_{AB}$ does not appear in the charges and captures (ingoing and memoryless) gravitational wave degrees of freedom that are part of the bulk phase space.

Before moving on we  mention a similar redundancy that is also present in the near bifurcate horizon analysis of section \ref{sec:3}, since the charges \eqref{charge-1-NB-case} depend only on the function $\Omega$ and the combination $(C_A^+ - C_A^-)/\eta$, as evident from \eqref{charge-densities-NB-case-1}. The phase space is parametrized by four functions $\eta$, $\Omega$ and $C^\pm_A$, and therefore there are two redundant functions. The redundancy is eliminated by gauge fixing, e.g.~$\eta=1$ and $C_A^-=0$.\footnote{Note that, given $\xi_{\tiny{{\cal P}}}:=\frac12 T(x^A)(x^+\partial_+-x^-\partial_-)$ and $\xi_{\tiny{{\cal D}}}:=\tilde T(x^A)(x^+\partial_++x^-\partial_-)$,  ${\cal P}$ is the charge associated with $\xi_{\tiny{{\cal P}}}$ and the charge for $\xi_{\tiny{{\cal D}}}$ vanishes on our phase space. That is, $\xi_{\tiny{{\cal D}}}$ generates trivial transformations which can be used to gauge fix $\eta$ to 1.}

\subsection{On-shell phase space for generic null hypersurfaces}\label{sec:6.2}

At first sight there seems to be a mismatch between the degrees of freedom of the boundary phase space and the number of surface charges since there are more functions on the phase space,  $\eta, \GH, \Omega, \Upsilon_A$, than {\NHS} generators, $T^\pm, Y^A$. The reason for this is that we did not consider all on-shell conditions yet. Therefore, we analyze now the equations of motion (eom) \eqref{EOM++Omega} and \eqref{EOM-C-A} and evaluate the charge and the flux \eqref{QI-MB} and \eqref{F-MB} on-shell.

The eom yield a dichotomy: generic solutions have $\partial_+\Omega\neq 0$, while special solutions have a non-expanding null hypersurface, $\partial_+\Omega=0$. We postpone a discussion of the latter case to section \ref{sec:6.3} and focus here on generic solutions.

In the generic case the eom \eqref{EOM++Omega} can be solved for $\GH$,
\begin{equation}\label{G-phase-space}
\GH=\GH(\eta,\Omega)= \partial_+ \left(\ln{\frac{\eta\,\Omega}{(\partial_+\Omega)^2}}\right)\,.
\end{equation}
Then the eom \eqref{EOM-C-A} yields
%\begin{equation}\label{Upsilon-phase-space}
%    \partial_+{\Upsilon_A}= 2\,\Omega\  \partial_+\partial_A \left(\ln{\frac{\partial_+\Omega}{\sqrt{\eta}}}\right)-\partial_+\Omega \frac{\partial_A\eta}{\eta}
%\end{equation}
%and is solved by $\Upsilon_A=\Upsilon_A(\eta,\Omega;\Upsilon_A^0)$, where $\Upsilon_A^0:=\Upsilon_A(x^+=0)$ is the initial value for $\Upsilon_A$. 
\begin{equation}\label{Upsilon-phase-space}
\partial_+ \Upsilon_A= 2\partial_+ \Big(\Omega\  \partial_A \Big(\ln{\frac{\partial_+\Omega}{\sqrt{\eta}}}\Big)-\partial_A\Omega\Big)\,.
\end{equation}
So $\Upsilon_A$ is fully determined by $\eta,\Omega$ and the initial value $\Upsilon_A^0:=\Upsilon_A(x^+=0)$.
%where $\Upsilon_A^0=\Upsilon_A^0(x^B)$. %So $\Upsilon_A$ is fully determined by $\eta,\Omega$ and $\Upsilon_A^0$ or alternatively by $\eta,\Omega$ and $\Upsilon_A(x^+=0)$.

Thus, generic points in the phase space (modulo possible `accidental gauge redundancies' which we discuss below) are uniquely specified on-shell by two arbitrary functions of time $x^+$, namely $\Omega$ and $\eta$, together with the initial value $\Upsilon_A^0(x^B)$. These on-shell degrees of freedom match the symmetry generating functions appearing in the near horizon Killing vectors, $T^\pm(x^+, x^A)$ and $Y^A(x^B)$, and the associated charges 
\begin{equation}\label{on-shell-charges-sec-5.5}
\begin{split}
{\cal Q}_\xi^{\eom} = &{\cal Q}_\xi^{\eom}[\Omega,\eta, \Upsilon_A^0; T^\pm, Y^A] \\ 
 := &(16\pi G)\left[\tilde{T}^- {\cal P}_- + Y^A {\cal J}_A +T^+ {\cal P}_+\right]
\end{split}
\end{equation}
where  $\tilde{T}^-:=\frac12(T^- +\partial_+T^+ )$. The charge densities ${\cal P}_\pm, {\cal J}_A$ read
\begin{subequations}\begin{align}
 {\cal J}_A&=\frac{\Upsilon_A}{16\pi G}\\
 {\mathbb {\cal P}}_- &=\frac{\Omega}{8\pi G}\\
 {\cal P}_+&=\frac{\Omega}{8\pi G}\partial_+
\left[ \ln{\frac{\partial_{+}\Omega}{\Omega}}- \ln {\sqrt{\eta\Omega}} \right]
\end{align}\end{subequations}
 and their variations yield~\footnote{%
 Note that ${\cal P}_\pm$ are scalar densities and ${\cal J}_A$ is a vector density. Therefore, $Y^A \partial_A {\cal P}_\pm+ {\cal P}_\pm\bar{\nabla}_AY^A=\bar{\nabla}_A \left(Y^A\,{\cal P}_\pm\right)$, where $\bar{\nabla}_A$ is the covariant derivative with respect to metric $\gamma_{AB}$.} 
\begin{subequations}\label{charge-variations-onshell}\begin{align}
 \delta_\xi{\cal J}_A&={\cal P}_- \partial_A \tilde T^- + T^+\partial_+ {\cal J}_A+{\cal P}_+\partial_A T^++\frac12{\partial_+} {\cal P}_-\partial_A T^+ \nonumber \\
 &\quad + Y^B\partial_B \mathcal{J}_A + \mathcal{J}_B\partial_A Y^B  + \mathcal{J}_A \partial_B Y^B\\
 \delta_\xi{ {\cal P}}_- &=T^+\partial_+ {\cal P}_-+Y^A \partial_A {\cal P}_-+ {\cal P}_-\bar{\nabla}_AY^A\\
 \delta_\xi{\cal P}_+&=   \partial_+\tilde T^- {\cal P}_- + \partial_{+} \left( T^+ {\cal P}_+\right)+ Y^A \partial_A {\cal P}_+ + {\cal P}_+ \bar{\nabla}_A Y^A\,.
\end{align}\end{subequations}
Thus, there are exactly as many charges, ${\cal P}_\pm, {\cal J}_A$, as there are independent functions on the boundary phase space,  $\Omega, \eta, \Upsilon^0_A$. As \eqref{charge-variations-onshell} show, the symmetry algebra of the on-shell charges is isomorphic to the symmetry algebra of the off-shell charges \eqref{eq:keyresult}.

While the charges ${\cal J}_A, {\cal P}_-$ are non-zero even in the limit of vanishing expansion, this is not true for ${\cal P}_+$. A reason for this significant difference is that the two former are integrable and conserved while the latter is not. This may be seen from the fact that the flux \eqref{F-MB} is proportional to $T^+$ and is independent of $T^-, Y^A$. Explicitly, the on-shell value of the flux is
\begin{equation}\label{on-shell-flux-sec-5.5}
{\cal F}_\xi^{\eom} = {\cal F}_\xi^{\eom}[\Omega,\eta; T^+] 
=2T^+\Big[\partial_+\big(\Omega \delta \ln \sqrt{\eta\Omega}\big) -\Omega \partial_+ \big(\delta \ln (\partial_+\Omega)\big) \Big]\,.
\end{equation}
The expression for the on-shell flux \eqref{on-shell-flux-sec-5.5}  is a function over the same on-shell phase space as the charges, and it is related to the latter through our GCCE \eqref{GCCE}.

\subsection{On-shell phase space for non-expanding {and} stationary null hypersurfaces}\label{sec:6.3}

We consider now non-expanding null hypersurfaces, $\partial_+\Omega=0$. This is a consistent truncation of the phase space since the condition $\partial_+\Omega=0$ is preserved under the orbit of the near null hypersurface Killing vectors.

In this case $\GH$ may not be solved for since the eom \eqref{EOM++Omega} is satisfied identically. The eom \eqref{EOM-C-A} and the expressions for charge \eqref{QI-MB} and flux \eqref{F-MB} simplify to
\begin{subequations}\label{QF-EOM-sec:6.2}
         \begin{align}
\label{Upsilon-stationary-phase-space}
    \partial_+{\Upsilon_A} +&\, \Omega\, \partial_A  \GH =0\\
 \label{QI-stationary}
  {\mathcal{Q}}^{\text{\tiny{non-ex}}}_{\xi} =& \, %\frac{1}{16\pi G}
2\,\tilde{T}^-\, \Omega  +Y^{A}\Upsilon_A -T^+\, \Omega\,\GH \\
 \label{F-stationary}
{\cal{F}}^{\text{\tiny{non-ex}}}_{\xi} (\delta g) =&\,  %\frac{T^{+}}{16 \pi G}
T^+ \Omega\, \delta \GH \,.
%= -\Omega\, \delta \mathbb{T} %\left({\cal G} -   \frac{\partial_+ \eta}{\eta}\right),
\end{align}
\end{subequations}

Within the special case of vanishing expansion we still encounter a dichotomy: generically, $\delta\GH\neq 0$, so despite of vanishing expansion there is a non-trivial flux, but for stationary solutions $\delta\GH=0$ and the flux vanishes. We discuss first the generic case.

\paragraph{Generic non-expanding case.} Our goal is to compare the number of free functions on the on-shell phase space with the number of free functions appearing in the surface charges. Since the functions ${\cal G}$ and $\eta$ always appear through the combination $\GH$ there is an accidental gauge symmetry, namely any transformation that changes $\eta, {\cal G}$ \eqref{VDF01-eta}, \eqref{VDF01-G} but keeps $\GH$ invariant. With no loss of generality we exploit this gauge symmetry to fix ${\cal G}=0$, implying 
\begin{equation}\label{GH-G=0-gauge}
    \GH=-\partial_+ \ln \eta\qquad \qquad \partial_+\Big(\Upsilon_A-\Omega\,\frac{\partial_A\eta}{\eta}\Big)=0
\end{equation}
where in the right equation we used \eqref{QI-stationary}. For consistency also its variation \eqref{VDF01-G} must vanish, yielding $\partial_+ T^-=0$ so that  $T^-= T^-(x^A)$, where we take $T^-$ to be field independent. 
The remaining eom \eqref{Upsilon-stationary-phase-space} implies that the on-shell degrees of freedom are $\GH (x^+,x^A)$, $\Omega(x^A)$ and $\Upsilon_A^0(x^B)$.  This matches precisely with the freedom contained in the symmetry generators $T^+(x^+, x^A)$, $T^- (x^A)$ and $Y^A(x^B)$. 

For completeness, we also present the on-shell variation of the remaining fields in this gauge from which one can read the algebra of charges. These variations may be obtained by setting $\partial_+\Omega=0$ and using $\partial_+T^-=0$ in  \eqref{delta-Omega}, \eqref{Upsilon-variation} and \eqref{T-C-G-variation},
\begin{subequations}\label{Field-Variations-Stationary}
\begin{align}
    \delta_\xi\Omega &=  Y^A \partial_A\Omega+\Omega\,\bar{\nabla}_A Y^A \label{deltaOmega-sec:6.2}\\
    \delta_\xi \Upsilon_A     & = \Omega\, \partial_A {T}^- +\Omega \partial_A (\partial_+ T^+ -T^+ \GH)     + Y^B\partial_B \Upsilon_A  + \Upsilon_B \partial_A Y^B + \Upsilon_A\partial_B Y^B \label{deltaUpsilon-sec:6.2}\\
  \delta_\xi  \GH &=  \partial_+ (T^+ \GH) -\partial^2_+ T^+ + Y^A\partial_A \GH\,. %= \partial_+ (T^+ \GH -\frac{Y^A}{\Omega} \Upsilon_A) 
  \label{deltaXi-sec:6.2}
\end{align}
\end{subequations}
In $\delta_\xi \Upsilon_A$ we dropped the term $T^+\left(\partial_+\Upsilon_A+\Omega \partial_A \ \GH\right)$, which vanishes on-shell \eqref{Upsilon-stationary-phase-space}.
In the ${\cal G}=0$ gauge the variation \eqref{deltaXi-sec:6.2} is equivalent to \eqref{VDF01-eta} for $\partial_+T^-=0$. For this sector we have the charges $R^{p,q}_n$, $T^{p,q}_0$, $L^p$ and ${\bar L}^r$. The T-Witt  part of the algebra hence reduces to a Witt algebra among $R^{p,q}_n$ and $T^{p,q}_0$ generators which commute with all the $R^{r,s}_n$. The generators $T^{p,q}_0$, $L^p$ and ${\bar L}^r$ form an algebra obtained in \cite{Donnay:2015abr} for the near horizon Kerr geometry.

\paragraph{Stationary case.} Finally, for the special case of stationary non-expanding null hypersurfaces, $\delta\GH=0$, the flux \eqref{F-stationary} vanishes and the charge \eqref{QI-stationary} is integrable
\begin{equation}\label{deltaQ-sec:6.2}
    {{Q}}^{\text{\tiny{stat}}}_{\xi} = \int_\gamma \big(\hat{T}\, \Omega + Y^A\, \Upsilon_A\big)\qquad \qquad \mathcal F_\xi=0
\end{equation}
where $\hat T:=T^- + \partial_+T^+ - \GH\,T^+$. The value of the charge and the charge variation has no $\eta$-dependence (or equivalently $\GH$-dependence). It implies that there is a new accidental gauge symmetry in the stationary sector. One can use this to gauge-fix $\eta$ to a constant and thus $\GH=0$. In this gauge $\hat T=T^- + \partial_+T^+$ and \eqref{deltaXi-sec:6.2} implies $\partial_+^2T^+=0$. Therefore, $\partial_+\hat T=\partial_+(T^- + \partial_+T^+)=0$. Moreover,
as a consequence of the eom \eqref{GH-G=0-gauge}, $\partial_+\Upsilon_A=0$. The charge \eqref{deltaQ-sec:6.2} is time independent and thus conserved, consistently with the GCCE \eqref{GCCE}.

Thus, also the stationary subsector of the boundary phase space shows a matching between charges, $\Omega(x^A)$ and $\Upsilon_A(x^B)$, and symmetry generators, $\hat T(x^A)$ and $Y^A(x^B)$. For the stationary phase space, $\GH=\partial_+\Omega=0$, the charge variation simplifies to
\begin{subequations}
\begin{align}
    \delta_\xi\Omega &=  Y^A \partial_A\Omega+\Omega\,\bar{\nabla}_A Y^A \label{deltaOmega-sec:6.2-2}\\
    \delta_\xi \Upsilon_A     & =2\, \Omega\, \partial_A {\hat T} + Y^B\partial_B \Upsilon_A  + \Upsilon_B \partial_A Y^B + \Upsilon_A\partial_B Y^B \, . 
\end{align}
\end{subequations}
The above is precisely what we obtained in section \ref{sec:3} as the symmetry algebra near the bifurcation point as well as the algebra obtained in \cite{Grumiller:2019fmp}. See also section \ref{sec:7.3} for further discussions.

We come back to bulk aspects of the phase space and gravitational wave excitations in the concluding section \ref{sec:8}. 

%#####################################################
\section{Recovering other near horizon symmetries}\label{sec:7}
%#####################################################

In this section we curtail the analysis of sections \ref{sec:4}, \ref{sec:5} and \ref{sec:onshellphasespace} by imposing physically well-motivated conditions on the general variations around the null hypersurface $x^-=0$. Thereby, we recover near horizon symmetries discussed earlier as special cases.

In section \ref{sec:7.1} we restrict the null hypersurface generator to have a constant non-affinity parameter over spacetime and phase space. In section \ref{sec:7.2} we 
discuss the Gaussian null coordinates. In section \ref{sec:7.3} we require the null hypersurface to be a  Killing horizon, which in a special case recovers the phase space established in section \ref{sec:3}.

\subsection{Null hypersurfaces with constant non-affinity parameter}\label{sec:7.1}

The generator of the $x^-=0$ null hypersurface in the metric \eqref{NFH-gauge-fixed} to leading order in $x^-$ is 
\begin{equation}\label{l-surface-grav}
l=\frac{N}{\eta}\, \partial_{+} +\mathcal{O}(x^-)
\end{equation}
where $N$ is an arbitrary function of $x^+$ and $x^A$. Its non-affinity parameter $\kappa_l$ defined through $l \cdot \nabla l^\mu=\kappa_l l^\mu$ on the surface $x^-=0$ is
\begin{equation}\label{surfgravhorgen}
    \kappa_{l}= \frac{1}{\eta}\left(\partial_{+} N -\frac{\mathcal{G} N}{2}\right)\,.
\end{equation}
The subscript $l$ is a reminder that the non-affinity parameter $\kappa_l$ in general is not associated with surface gravity $\kappa$, since we are currently not assuming to have a Killing horizon; for the special case of a Killing horizon both quantities coincide with each other.

Requiring $\kappa_l$ to be constant on both spacetime and phase space, i.e., $\partial_\mu \kappa_l=0$ and $ \delta \kappa_l=0$, fixes $N$ as a function of ${\cal G}$ and $\eta$. This does not impose any further restriction on our phase space. However, physically motivated choices can lead to further restrictions. Here we focus on the case $N\propto x^+\eta$. Constancy of $\kappa_l$ in this case leads to
\begin{equation}\label{kappa-l-constant-GNC}
\mathcal G=2\frac{\partial_+\eta}{\eta}\,.
\end{equation}
The above condition constitutes a special case for which the $\eta,{\cal G}$-dependence drops out of the field equation \eqref{EOM++Omega}, yielding
\begin{equation}\label{Omega-GNC}
    \Omega= \Omega_0+2\sqrt{\Omega_{0}\Omega_1} \ x^+ +\Omega_1 \ (x^+)^2\qquad\qquad \Omega_{0}=\Omega_{0}(x^A),\ \Omega_{1}=\Omega_{1}(x^A)\,.
\end{equation}
Recalling the variations \eqref{VDF01}, to preserve the condition \eqref{kappa-l-constant-GNC} on the phase space the transformation parameter $T^+$ has to obey
\begin{equation}\label{kappa-l-constant-T-'}
  \partial^2_+T^+=0 \qquad \Longrightarrow\qquad  T^+=T_0+x^+ T_1
\end{equation}
where $T_0, T_1$ are arbitrary functions of $x^A$, while $T^-$ and $Y^A$ are unrestricted. Therefore, the on-shell phase space in general is determined by $\eta(x^+,x^A), \Omega_0(x^A), \Omega_1(x^A), \Upsilon_A^0(x^B)$ and the symmetry generators are 
$T^-(x^+,x^A), T_0(x^A), T_1(x^A), Y^A(x^B)$.

\subsection{Null hypersurfaces in Gaussian null coordinates}\label{sec:7.2}

Many previous near horizon studies invoked Gaussian null coordinates, see e.g.~\cite{Donnay:2015abr,Donnay:2016ejv,Chandrasekaran:2018aop,Chandrasekaran:2019ewn}. It is one of the purposes of this subsection to show that there is loss of generality in assuming Gaussian null coordinates, because the coordinate transformations required to transform to these coordinates are not proper ones within our more  general set of boundary conditions \eqref{metric-gauge-fixed-sec:4}. 
Technically, it turns out that Gaussian null coordinates require to fix $\eta=1$ and $\delta\eta=0$ on the phase space of metrics \eqref{metric-gauge-fixed-sec:4}. This leads to a reduced phase space. Hence there is loss of generality in making these assumptions. In the remainder of this subsection we make explicit the consequences of the choice $\eta=1$ (and $\delta\eta=0$), in particular for the metric and for the symmetry generators. 

The condition $\eta=1$ is achieved as follows. Consider the coordinate system in section \ref{sec:5.4}, in particular the  metric \eqref{metric-gauge-fixed-sec:4}. Next, perform the coordinate transformation %\eqref{AKVH01}
\begin{equation}\label{Bx-'}
    x^- \to x^-  \mathcal{C} 
\end{equation}
where $\mathcal{C}$ is an arbitrary function of $x^+$ and $x^A$. The functions in the metric \eqref{metric-gauge-fixed-sec:4} then transform as
\begin{equation}\label{eq:C-transf}
  \eta\to  {\cal C}\eta\qquad\qquad {{\cal G}} \to {{\cal G}}+2\,\frac{\partial_+{\cal C}}{{\cal C}}\qquad\qquad \frac{1}{\eta}\theta_A\to \frac{1}{\eta} \theta_A+\frac{\partial_A{\cal C}}{{\cal C}}\,.
\end{equation}
The coordinate transformation \eqref{Bx-'} is generated by $T^-$ in \eqref{AKVH01}. The transformations of the metric functions \eqref{eq:C-transf} leave invariant $\Omega$ and ${\cal G}-2\partial_+\eta/\eta$, while ${\cal G}$, $\eta$, and $\Upsilon_A$ transform. Therefore, the coordinate transformation \eqref{Bx-'} is a nontrivial diffeomorphism since there is a non-vanishing surface charge associated with it. The choice ${\cal C}=1/\eta$ yields the desired reduction of the phase space to a sector in which $\eta=1$. The metric \eqref{metric-gauge-fixed-sec:4} is then manifestly  in the Gaussian null coordinates,
\begin{equation}\label{metric-gauge-fixed-sec:5-2-2'}
\d s^2=-2x^-{\cal K}\ (\d x^+){}^2+2\d x^- \d x^+ +2x^- \tilde\theta_A \d x^+\d x^A +\Omega_{AB}\d x^A\d x^B+\cdots
\end{equation}
where
\begin{equation}\label{tilde-Kappa'}
    {\cal K}= -\frac{\mathcal{G}}{2}+ \frac{\partial_+\eta}{\eta}\qquad
    \qquad \tilde\theta_A=\frac{1}{\eta}(\theta_A-\partial_A\eta)\,.
\end{equation}

Due to the form of the near null hypersurface Killing vectors \eqref{AKVH01} a generic transformation moves us away from $\eta=1$. To keep this condition we require $\delta_\xi\eta=0$, which by virtue of \eqref{VDF01-eta} yields
\begin{equation}\label{Tpm-sec-6.2.2'}
    T^-=\partial_+ T^+\,.
\end{equation}
One may use \eqref{EOM++Omega} to solve ${\cal G}$  in terms of $\Omega$ and then \eqref{EOM-C-A} to solve $\Upsilon_A$ in terms of $\Omega$ and an initial value function $\Upsilon_A^0$. So in this sector our on-shell phase space is described by $\Omega(x^+, x^A), \Upsilon_A^0(x^B)$ and the independent symmetry generators are $T^+(x^+, x^A), Y^A(x^B)$. %(Note that ${\cal K}$ through field equation \eqref{EOM++Omega} may be written as a function of $\Omega$.) 
The T-Witt algebra \eqref{Future-horizon-algebra-of-generators-BMS3} is then reduced to the S-Witt algebra, generated by ${\cal R}_n^{p,q}$.

In addition to the case $\eta=1$ considered here, one can demand the generator $l^\mu\partial_\mu = N \, \partial_++\mathcal O(x^-)$ to have a constant non-affinity parameter, as it was done in section \ref{sec:7.1}. The constant non-affinity parameter case with $N\propto x^+\eta$ yields \eqref{kappa-l-constant-GNC} and does not constrain $T^-$. It is hence compatible with the $\eta=1$ constraint and the two restrictions may be combined. 

Requiring constancy of the non-affinity parameter in \eqref{kappa-l-constant-GNC} together with $\eta=1$ implies ${\cal G}=0$. Preserving this condition yields
\begin{equation}\label{Tpm-GNC}
  \partial^2_+T^+=0,  \  T^-=\partial_+ T^+ \qquad \Longrightarrow\qquad  T^+=T_0+x^+ T_1\,,\ T^-=T_1
\end{equation}
where $T_0, T_1$ are arbitrary functions of $x^A$. The field equations for $\Omega$ yield \eqref{Omega-GNC} and for $\Upsilon_A$ \eqref{EOM-C-A} reduce to $\partial_+ \Upsilon_A=\Omega \partial_+\partial_A \ln \Omega$. Therefore, time dependence of $\Upsilon_A$ is completely determined in terms of $\Omega$, the on-shell phase space in general is specified by $\Omega_0(x^A), \Omega_1(x^A), \Upsilon_A^0(x^B)$, and the symmetry generators are 
$T_0(x^A), T_1(x^A), Y^A(x^B)$. The symmetry generators, charge and flux read
\begin{subequations}
\label{xi-Q-F-GNC}
\begin{align}
     \xi&= T_0 \partial_+ + T_1 (x^+\partial_+- x^-\partial_-) + Y^A \partial_A\\
        {\cal Q}_\xi &=2\Omega\partial_+ T^+  -2 T^+\partial_+ \Omega + Y^A \Upsilon_A\\
        {\cal F}_\xi &={4\,}T^+\sqrt{\Omega_1}\ \delta(\sqrt{\Omega})\,. 
\end{align}
\end{subequations}
The charge becomes integrable when the flux vanishes, i.e., for $\Omega_1=0$. This is the stationary  case with $\partial_+\Omega=0$, discussed in section \ref{sec:6.3}.

We finally note that upon the coordinate transformation
\begin{equation}\label{xpm-vrho'}
     x^+ =\  e^{{\kappa} v}
     + \mathcal{O}(\rho^2)\qquad \qquad
    x^- =\  \frac{e^{-{\kappa} v}}{\kappa} \rho+\mathcal{O}(\rho^2)\qquad  \qquad
    x^A =\  y^A
    + \mathcal{O}(\rho^2)\,,
\end{equation}
with $\kappa:=\kappa_l$, the line-element \eqref{metric-gauge-fixed-sec:4} (for $\eta=1, {\cal G}=0$) takes the standard Gaussian null form
\begin{equation}\label{EF-type-01'}
    \d s^2=-2\,{\kappa}\,\rho\,\d v^2+2\,\d v\,\d \rho+2\,\rho\,{\theta}_A\, \d v\,\d y^A+\left({\Omega}_{AB}+\rho\,{\lambda}_{AB}\right)\,\d y^A\,\d y^B+\mathcal{O}\left(\rho^2\right)\,.
\end{equation}   
The null hypersurface $x^-=0$ is mapped to the surface $\rho=0$ in \eqref{EF-type-01'} and is generated by 
$\zeta_{\textrm{\tiny H}}= \kappa\, ( x^+\partial_+-x^- \partial_-)=\partial_v$ whose the non-affinity parameter $\kappa$ is constant. Note that the coordinate transformation \eqref{xpm-vrho'} is a proper gauge transformation over the restricted phase space we are considering in this section. 

The near horizon Killing vector \eqref{xi-Q-F-GNC} in Gaussian null coordinates 
\begin{equation}\label{DGGP-Vector'}
    \xi = f \partial_{v} - \partial_{v}f \rho \partial_{\rho} + \left( Y^A - \rho \, {\Omega}^{AB} \partial_{B} f\right) \partial_{A} + \mathcal{O}(\rho^2) \quad\textrm{with}\quad f= \frac1\kappa\,\big(T_1+ e^{-\kappa v} T_0\big)
\end{equation}
is exactly the near-horizon symmetry generator for isolated horizons presented in \cite{Donnay:2016ejv, Chandrasekaran:2019ewn}.  The near horizon symmetry algebra in this case is given by \eqref{eq:angelinajolie}, but with the T-Witt sector restricted to ${\cal R}^{p,q}_0\simeq {\cal R}^{p,q}_0+\mathcal{T}^{p,q}_{0}$ (which in terms of differential operators is $-\zeta_{\textrm{\tiny H}}/\kappa$) and ${\cal R}^{p,q}_{-1}$.

%%%%%%%%%%%%%%%%%%%%%%%%%%%%%%%%%%%%%%%%%%%%%%%%%%%%%%%%%%%%%%%%%%%%%%
\subsection{Killing horizons}\label{sec:7.3}
%%%%%%%%%%%%%%%%%%%%%%%%%%%%%%%%%%%%%%%%%%%%%%%%%%%%%%%%%%%%%%%%%%%%%%
In this subsection we restrict the phase space to describe Killing horizons. It means we require the existence of a vector field $\zeta_{\text{\tiny H}}$ normal to the null hypersurface $x^-=0$  such that ${\cal L}_{\zeta_{\text{\tiny H}}} g_{\mu\nu}=0$ and $\zeta_{\text{\tiny H}}^2=0$ at  $x^-=0$. The geometry \eqref{metric-gauge-fixed-sec:4} admits such a  Killing vector field, 
\begin{equation}\label{genhor}
   \zeta_{\text{\tiny H}} =\frac{N}{\eta}\partial_++l_{(1)}^\mu\,x^-\,\partial_\mu+\mathcal O(x^-)^2
\end{equation}
where $N$ is a possibly field-dependent function over the phase space, if 
\begin{subequations}\label{eqgenhortheta}
\begin{align}
    \partial_+\Omega&=0 &  \GH&= \partial_+\ln \frac{N^2}{\eta}-2\frac{\eta}{N} \kappa &&
    \label{6.19a}\\
    \Omega_{AB}l_{(1)}^B&=-N \partial_A\ln\frac{N}{\eta} &
    l_{(1)}^-&=-\frac{\partial_+N}\eta &
    l_{(1)}^+&=0\label{6.19b}
    \\ \partial_+{\Upsilon_A} + \Omega\, \partial_A  \GH&=\frac{2 \eta}{N}  \partial_{A}\kappa &&&& \label{6.19c}
\end{align}
\end{subequations}
where $\kappa(x^A)$ is the non affinity parameter of \eqref{genhor}. Moreover the eom $\Upsilon_A$ \eqref{Upsilon-stationary-phase-space} implies $\partial_A \kappa=0$. This is the zeroth law of black hole thermodynamics stating that the surface gravity $\kappa$ has to be constant over the Killing horizon. 

The Killing horizon case is therefore a special case of the non-expanding case of section \ref{sec:6.3} which has constant surface gravity. The charges and their algebra are hence the same as the one discussed below \eqref{Field-Variations-Stationary}.  
As in section \ref{sec:6.3}, one may fix ${\cal G}=0$ gauge in which $\partial_+N=\kappa \eta$, and thus $N=\kappa\, \int^{x^+} \eta+U(x^A)$. However the integration function $U(x^A)$ has to vanish in order to consistently describe the bifurcation point. The variation of $N$ is 
\begin{equation}\label{delta-N}
    \delta_{\xi} {N}= T^+\partial_{+} N- T^- N  +Y^A \partial_{A}{ N}%+  U(x^A)
\end{equation}
where we used \eqref{VDF01-eta} and the fact that  $T^-=T^-(x^A)$. 

As in the previous subsection, one may further restrict $N$. For example, for the case of $N=\kappa \eta x^+$, $\partial_+N=\kappa \eta$ yields $\partial_+\eta=0$ and \eqref{6.19b} yields $l_{(1)}^A=0$. Therefore, 
\begin{equation}\label{killingvector5}
\zeta_{\text{\tiny H}} =  \kappa \left( x^+\partial_+ -x^- \partial_{-} \right) +\mathcal O(x^-)^2 \,.
\end{equation} 
Moreover, in this case $\Gamma=0, \delta\Gamma=0$ and hence we recover the stationary case of  section \ref{sec:6.3}. That is, after fixing the extra accidental symmetries we find in this case, the on-shell physical phase space is described by $\Omega(x^A), \Upsilon_A^0(x^B)$ and the symmetry algebra reduces to that of section \ref{sec:3}.

%%%%%%%%%%%%%%%%%%%%%%%%%%%%%%%%%%%%%%%%%%%%%%%%%%%%%%%%%%%%%5
\section{Concluding remarks}\label{sec:8}
%%%%%%%%%%%%%%%%%%%%%%%%%%%%%%%%%%%%%%%%%%%%%%%%%%%%%%%%%%%%%%'

We conclude with a brief summary of the main results and a perspective on future research directions. 

Motivated by the desire to understand non-extremal Kerr black holes from a near horizon perspective we introduced new coordinates \eqref{eq:kerr1} that are co-rotating with the event horizon but otherwise do not mix angular and spacetime coordinates and allow to cover the whole Kerr manifold up to the Cauchy horizon. Expanding around the bifurcation 2-sphere of the Kerr geometry led us to propose bifurcation fall-off conditions \eqref{FB01}. The associated bifurcation Killing vector algebra \eqref{NHA-B} consisted of superrotations and two towers of supertranslations. However, only one such copy entered the surface charges \eqref{charge-1-NB-case}, while as we discussed in the end of section \ref{sec:5.4}, the other one turned out to be pure gauge. 

We generalized the fall-off conditions to generic (but co-rotating) null hypersurfaces (regardless of whether they describe bifurcate Killing horizons, isolated horizons or no horizons), see \eqref{NFH01}. The near null hypersurface Killing vectors \eqref{AKVH01} generated an algebra \eqref{eq:angelinajolie} that contained again superrotations and two towers of supertranslations, which among themselves formed a subalgebra \eqref{Future-horizon-algebra-of-generators-BMS3} that we dubbed `T-Witt' algebra. In contrast  to the previous case both towers of supertranslations turned out to be relevant for the charges \eqref{Non-I-Ch}, which no longer were integrable. Their split into integrable and flux parts was ambiguous, as expected on general grounds, but the ambiguity could be largely fixed by demanding the absence of a central extension. This led us to the integrable part \eqref{QI-MB} and the flux part \eqref{F-MB}, which together obey a generalized charge conservation equation \eqref{GCCE} [or \eqref{entropy-flux}] that justifies the label `flux'. We verified in section \ref{sec:onshellphasespace} that there are always equally many functions parametrizing the on-shell phase space and functions appearing in the symmetry generators, which is why we referred to the phase space as `boundary phase space'. In section \ref{sec:7} we recovered special cases considered in earlier literature on near horizon symmetries.

For readers exclusively interested in results pertaining to Kerr black holes \eqref{eq:kerr1} we summarize them briefly in this paragraph. The near bifurcation charges and their variations are displayed in \eqref{eq:kerrcharges}-\eqref{charge-variation-NB-case-P-basis}. The more generic near horizon expansion of Kerr is displayed in appendix \ref{Append:Kerr-expansion}.

%%%%%%%%%%%%%%%%%%%%%
%
% FIG 1 STARTS BELOW
%
%%%%%%%%%%%%%%%%%%%%% 
% .--..--..--..--.
%/ .. \.. \.. \.. \
%\ \/\ \/\ \/\ \/ /
% \/ /\/ /\/ /\/ /
% / /\/ /\/ /\/ /\
%/ /\ \/\ \/\ \/\ \
%\ \/\ \/\ \/\ \/ /
% \/ /\/ /\/ /\/ /
% / /\/ /\/ /\/ /\
%/ /\ \/\ \/\ \/\ \
%\ \/\ \/\ \/\ \/ /
% \/ /\/ /\/ /\/ /
% / /\/ /\/ /\/ /\
%/ /\ \/\ \/\ \/\ \
%\ `'\ `'\ `'\ `' /
% `--'`--'`--'`--'
%%%%%%%%%%%%%%%%%%%%%
\begin{figure}[ht]
\bigskip
\begin{center}
% Define block styles
\tikzstyle{decision} = [diamond, draw, fill=blue!30, text width=6em, text badly centered, node distance=3cm, inner sep=2pt]
\tikzstyle{block} = [rectangle, draw, fill=taplum!90, text width=8em, text centered, minimum height=4em]
\tikzstyle{line} = [draw, -latex']
\tikzstyle{endblock} = [rectangle, draw, fill=taorange, text width=8em, text centered, rounded corners, minimum height=4em]
\tikzstyle{inblock} = [rectangle, draw, fill=tachameleon!60, text width=8em, text centered, rounded corners, minimum height=4em, drop shadow]
\tikzstyle{solblock} = [rectangle, draw, fill=tabutter, text width=8em, text centered, rounded corners, minimum height=4em]
\begin{tikzpicture}[node distance = 3cm,auto]
    % Place nodes
    \node [inblock] (twitt) {generic:~T-Witt\\${\cal R}_n^{p,q}$, ${\cal T}_n^{p,q}$\\Eq.~\eqref{Future-horizon-algebra-of-generators-BMS3}};
    \node [solblock, below of=twitt, node distance=1cm, right of=twitt, node distance=3.5cm] (nonex) {non-expanding\\${\cal R}_n^{p,q}$, ${\cal T}_0^{p,q}$\\Sec.~\ref{sec:6.3}};
    \node [solblock, below of=nonex, node distance=2.5cm] (switt) {GNC:~S-Witt\\${\cal R}_n^{p,q}$\\Sec.~\ref{sec:7.2}};
    \node [solblock, below of=twitt, node distance=1cm, left of=twitt, node distance=3.5cm] (const) {const.~non-affinity\\${\cal R}_{-1}^{p,q}$, ${\cal R}_0^{p,q}$, ${\cal T}_n^{p,q}$\\Sec.~\ref{sec:7.1}};
    \node [endblock, below of=const, node distance=2.5cm] (gnc) {GNC \& const.~$\kappa$\\${\cal R}_{-1}^{p,q}$, ${\cal R}_0^{p,q}$\\Sec.~\ref{sec:7.2}};
    \node [endblock, below of=twitt, node distance=9.5cm] (bi) {stationary\\${\cal R}_0^{p,q}\simeq {\cal R}_0^{p,q}+{\cal T}_0^{p,q}$\\Sec.~\ref{sec:3} \& \ref{sec:6.3} \& \ref{sec:7.3}};
    \path [line] (twitt) -- (nonex);
    \path [line] (twitt) -- (const);
    \path [line] (nonex) -- (switt);
    \path [line] (switt) -- (gnc);
    \path [line] (const) -- (gnc);
    \path [line] (gnc) -- (bi);
    \path [line] (switt) -- (bi);
\end{tikzpicture}
\caption{Selected infinite subalgebras of T-Witt algebra. The  {\NHS} algebra consists of superrotations and the T-Witt. Some physically motivated restrictions, such as vanishing expansion, leave intact the superrotations but reduce T-Witt to one of its subalgebras discussed in appendix \ref{Appen:B}. Particularly the assumption of Gaussian null coordinates reduces T-Witt to S-Witt. Assuming additionally constant surface gravity and/or stationarity yields further reductions, recovering near horizon symmetries also found in previous literature. }
\label{fig:1}
\end{center}
\bigskip
\end{figure}
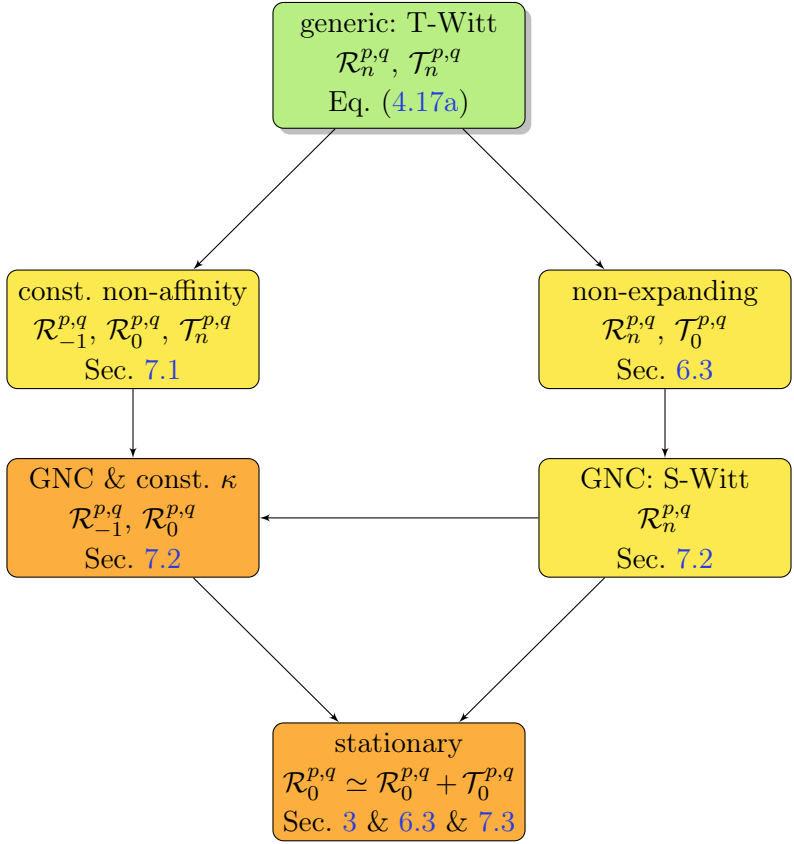
%%%%%%%%%%%%%%%
%
% END OF FIG 1
%
%%%%%%%%%%%%%%%

In this work we considered some subalgebras of the T-Witt algebra motivated by geometric and/or physical restrictions of the boundary conditions, see fig.~\ref{fig:1} and appendix \ref{Appen:B}. There are numerous further algebraic aspects that deserve further study. We list here some of them. It could be interesting to fully classify all maximal finite subalgebras and the infinite subalgebras of the T-Witt algebra, as well as all their central extensions. To obtain the latter we would need to relax the regularity assumption stated below \eqref{eq:angelinajolie} (e.g.~by compactifying the null hypersurface or by allowing a singular point $x^+ = 0$) and allow for positive and negative mode numbers. Besides these purely algebraic aspects, it would be physically interesting to find a geometric realizations of boundary conditions that switch on some of the algebraically allowed central extensions. Also the superrotation part of the algebra may deserve further study, see e.g.~\cite{Campiglia:2020qvc} and refs.~therein for a recent discussion in the context of asymptotically flat spacetimes.

Some current and future research directions that have strong overlap with the results of our work are listed below. 
\begin{itemize}
    \item {\bf Bulk and boundary.} The discussions in our work also reflect on purely theoretical considerations relevant for field theories in the presence of boundaries, see e.g.~\cite{Compere:2018aar,Strominger:2017zoo,Harlow:2019yfa,Geiller:2019bti,Barnich:2019xhd} and refs.~therein. On general grounds, the phase space splits into a bulk part, capturing the local physical excitations (in our case two helicities of massless gravitational waves), and a boundary part, capturing physical boundary degrees of freedom. While there seems universal agreement on the bulk, there are different perspectives in the literature on how to interpret the boundary degrees of freedom, sometimes even to the extent that it is questioned whether they are physical. A common perspective is to view the surface charges as a necessary boundary contribution to the canonical gauge generators to render them functionally differentiable. Following this line of thought to its conclusion means that the surface charges parametrize those boundary condition preserving transformations that are pure gauge in the bulk, but fail to be pure gauge at the boundary. Hence, there are fewer gauge degrees of freedom at the boundary, which is the essence of why physical boundary excitations emerge. An alternative perspective is to explicitly add boundary degrees of freedom with their own boundary equations of motion to restore gauge invariance  and factorizability of the Hilbert space under splitting of spacetime into subregions, see \cite{Geiller:2019bti} and refs.~therein. Either way, the total physical phase space consists of two parts, the bulk phase space (describing gravitational wave excitations in Einstein gravity) and the boundary phase space (described by surface charges and governed by the symmetries they generate). In our work we focused almost exclusively on the latter. It would be interesting to generalize our discussion to address also the bulk phase space (the gravitational waves) and its interaction with boundary degrees of freedom, for instance to describe physical processes such as  formation of a black hole, things falling into the hole or black hole evaporation and their impact on the surface charges.
    \item {\bf Gravitational waves.} As a first step to generalize our work to include a discussion of bulk degrees of freedom we can identify where gravitational wave excitations are hidden in our construction. In the coordinates adapted to null hypersurfaces we have an ingoing direction $x^-$, an outgoing direction $x^+$ and transversal directions $x^A$, see for instance the fully gauge-fixed version of metrics \eqref{metric-gauge-fixed-sec:4}. Ingoing gravitational waves can be Fourier-decomposed with factors $e^{ik_- x^-}$, and thus are captured by metric functions such as $\lambda_{AB}$ (and subleading terms) in  \eqref{metric-gauge-fixed-sec:4}. Outgoing gravitational waves with Fourier factors $e^{ik_+x^+}$ are switched off by the conformality assumption \eqref{2d-metric}, so by relaxing this assumption the information about such waves would be contained in $\Omega_{AB}$. Thus, we see that three quarters of all gravitational wave modes require to drop the assumption \eqref{2d-metric}; the remaining quarter, which is already included in our current work, are ingoing gravitational waves with nodes on the null hypersurface. The latter do not influence the surface charges, while the remaining three quarters will generically induce non-trivial changes of the surface charges and fluxes. It should be rewarding to consider some elementary processes --- e.g.~capture or emission of a single plane graviational wave --- to unravel the possible interactions between surface charges and gravitational waves, and more generally between boundary and bulk degrees of freedom. 
    \item {\bf Charges and fluxes.} The charges associated with boundary conditions in the present work are not generically integrable. The Barnich--Troessaert method of using a modified bracket almost uniquely allows to separate the integrable from the non-integrable part (there remains an inessential ambiguity, up to algebra 2-cocycles). The generalized charge conservation equation \eqref{GCCE} relates the non-integrability of the charge to the non-conservation of the integrable part. Instead of using the modified bracket one could elevate the generalized charge conservation equation to a postulate. This is essentially the idea behind the Wald--Zoupas method \cite{Wald:1999wa}: if there is a region of stationary configurations in the phase space one expects there the charges to be conserved (and hence integrable), and the fluxes to vanish. This region may then be used as reference point for charge-flux separation. In all examples that both the Barnich--Troessaert and Wald--Zoupas methods can be applied to they yield the same result. In our case, however, as the discussions in section \ref{sec:onshellphasespace} reveal, in general we do not have a reference point of stationary configurations. Hence, one may not  unambiguously employ the Wald--Zoupas method. It could be interesting to explore if the Wald--Zoupas method can be generalized to cover cases like ours.
    \item {\bf Soft hair.} Numerous previous papers considered near horizon aspects of black holes, see e.g.~\cite{tHooft:1990fkf,tHooft:1991uqr,Susskind:1993if,Hayward:1993wb,Carlip:1994gy,Balachandran:1994up,Carlip:1995cd,Strominger:1997eq,Ashtekar:1997yu,Carlip:1998wz,Hotta:2000gx,Ashtekar:2000sz,Ashtekar:2002ag,tHooft:2006xjp,Majhi:2012tf,Penna:2015gza,Afshar:2015wjm} and refs.~therein. A new aspect addressed during the past five years is the emergence of soft hair excitations. The name `soft hair' was coined by Hawking, Perry and Strominger \cite{Hawking:2016msc}, where `soft' refers to zero energy excitations and `hair' to these excitations being physical. This line of research has engendered a lot of recent activities, see for instance \cite{Hooft:2016itl,Averin:2016ybl,Compere:2016hzt,Afshar:2016wfy,Eling:2016xlx,Afshar:2016uax,Mirbabayi:2016axw,Donnay:2016ejv,Hopfmuller:2016scf,Hawking:2016sgy,Afshar:2016kjj,Wieland:2017zkf,Bousso:2017dny,Strominger:2017aeh,Akhmedov:2017ftb,Lust:2017gez,Blommaert:2018rsf,Hooft:2018gtw,Donnelly:2018nbv,Chandrasekaran:2018aop,Donnay:2018ckb,Haco:2018ske,Penna:2018gfx,Donnay:2019jiz,Wieland:2019hkz,Grumiller:2019tyl,Chandrasekaran:2019ewn,Grumiller:2019fmp,Grumiller:2019ygj,Bagchi:2019clu,Ashtekar:2020ifw}. The first papers where soft hair was constructed as near horizon excitations were based on boundary conditions that kept fixed surface gravity and had integrable surface charges \cite{Donnay:2015abr,Afshar:2016wfy}, unlike the present work. A subtlety in these constructions is the precise state-dependence of the near horizon Killing vectors: depending on the assumptions one can construct different near horizon symmetries,  with or without central charges, including BMS symmetries with supertranslation generators of arbitrary spin and Heisenberg symmetries \cite{Grumiller:2019fmp}. It is possible (and could be useful for applications) to generalize the construction in our present work to allow for suitable state-dependence of the near null hypersurface Killing vectors \eqref{AKVH01}, i.e., to allow certain field-variations of the functions $T^\pm$ and $Y^A$ appearing therein. Like in \cite{Grumiller:2019fmp} this would model the physical properties of the black hole and its interactions with some fiducial thermal bath, much like macroscopic electrodynamics models certain materials through suitable boundary conditions without having to deal with details of the microphysics. The novel aspect provided by our results in the context of this program compared to the discussions in \cite{Grumiller:2019fmp} is the possibility to allow for non-integrable charges and hence fluxes through the horizon. 
    \item {\bf Thermodynamical aspects.} In the stationary limit there is a simple near horizon first law relating variations of the near horizon energy to temperature times the variation of the black hole entropy \cite{Donnay:2015abr}, as expected from the Iyer--Wald derivation \cite{Iyer:1994ys} of the first law of black hole mechanics (see also \cite{Hajian:2015xlp} for generalizations of the Iyer--Wald analysis). For time-dependent situations considered in the present work it is plausible that at least in the adiabatic limit there could be a generalized first law relating time variations of the near horizon energy to time variations of entropy and fluxes across the horizon. It will be interesting to establish such a generalized first law, based on the generalized charge conservation equation \eqref{entropy-flux}.
    \item {\bf Kerr entropy.} One of the outstanding goals is to provide a microscopic picture of the entropy of generic Kerr black holes. While for the extremal case one may invoke the Kerr/CFT correspondence \cite{Guica:2008mu,Compere:2012jk} to count the black hole microstates, it remains unclear what these microstates are and how to generalize this construction to generic Kerr black holes. Haco, Hawking, Perry and Strominger \cite{Haco:2018ske} proposed the two-dimensional conformal algebra (with central charges given by the black hole angular momentum times 12) near the horizon of generic Kerr black holes as responsible for the microstates and verified that applying the Cardy formula to the conformal algebra reproduces the correct Bekenstein--Hawking entropy. A discussion by Aggarwal, Castro and Detournay \cite{Aggarwal:2019iay} suggests alternatively at the same level of rigor centrally extended Virasoro--Kac--Moody symmetries. In the present work as in many previous works, e.g.~\cite{Donnay:2015abr, Donnay:2016ejv, Chandrasekaran:2018aop, Chandrasekaran:2019ewn, Grumiller:2019fmp, Flanagan:2019vbl}, the near horizon symmetry algebras do not have central charges. Absence of central terms in the canonical realization of the near horizon symmetry algebra bars simple `Cardyology' but does not mean that these symmetry algebras cannot be used for microstate counting, as long as there is a gap between the vacuum and the black holes state. While we do not know if the T-Witt symmetries of our present work  will ultimately be useful for a generic Kerr black hole microstate counting, it is an exciting possibility to ponder about in the future. 
    \item {\bf Semi-classical microstate construction.} Even if the microstate counting based on symmetries did work, it does not tell us precisely what these microstates are. A more ambitious approach is to explicitly construct these microstates, at least in some semi-classical limit (large black holes, sufficiently far away from extremality). We have no idea if this ever will work for generic Kerr black holes, but we offered a concrete ``fluff'' proposal for three-dimensional black holes in anti-de~Sitter space \cite{Afshar:2016uax,Afshar:2017okz} that was based on near horizon soft hair. This construction gives at least a glimpse of hope that near horizon symmetries, such as the ones discussed in the present work, will be a key aspect in semi-classical microstate constructions of generic Kerr black holes. 
    \item {\bf Information loss.}  Charge non-conservation/non-integrability and generalized conservation equations like \eqref{GCCE} and \eqref{entropy-flux} may also be useful to move beyond stationary questions (like microstates) and address dynamical ones, e.g.~Hawking radiation and the information loss problem. It would be rewarding to explore this further.
\end{itemize}
Finally, the effect that assuming some a priori gauge like Gaussian null coordinates can reduce the boundary phase space is independent from whether one expands near the horizon or in the asymptotic region. Indeed, a similar effect was observed already in three-dimensional gravity where the assumption of Fefferman--Graham gauge reduces the physical phase space \cite{Grumiller:2016pqb,Grumiller:2017sjh}. Thus, it seems plausible that also in the context of four-dimensional asymptotically flat gravity standard assumptions like Bondi-gauge can reduce the boundary phase space. In conclusion, it would be interesting to reconsider yet-another-time asymptotic symmetries of asymptotically flat spacetimes to verify whether or not the same effect arises there. It cannot be excluded that an enlarged set of asymptotic symmetries will allow to make novel statements about the infrared triangle asymptotic symmetries/memory effects/soft theorems \cite{Strominger:2013jfa,Strominger:2017zoo,Pasterski:2019msg}.

\section*{Acknowledgement}

We are grateful to Glenn Barnich, Laura Donnay, Laurent Freidel, Marc Geiller, Gaston Giribet, Mehdi Hakami Shalamzari, Marc Henneaux, Malcolm Perry, Kartik Prabhu, Andrea Puhm, Romain Ruzziconi and Wolfgang Wieland for useful discussions. DG and MMShJ thank Hamid Afshar, Martin Ammon, Stephane Detournay, Hern\'an Gonz\'alez, Philip Hacker, Wout Merbis, Blagoje Oblak, Alfredo P\'erez, Stefan Prohazka, David Tempo, Ricardo Troncoso, Raphaela Wutte and Hossein Yavartanoo for previous collaboration on near horizon symmetries.

HA, SS and MMShJ acknowledge the support by INSF grant No.~950124 and Saramadan grant No.~ISEF/M/98204. 
DG was supported by the Austrian Science Fund (FWF), projects P~30822 and P~32581. 
CZ was supported by the Austrian Science Fund (FWF), projects P~30822 and M~2665. 
DG, SS and MMShJ acknowledge the Iran-Austria IMPULSE project grant, supported and run by Khawrizmi University and OeAD.
MMShJ would like to thank the hospitality of ICTP HECAP and ICTP EAIFR where a part of this work carried out. 

\appendix
%\addcontentsline{toc}{section}{Appendices}
\newcommand{\nocontentsline}[3]{}
\newcommand{\tocless}[2]{\bgroup\let\addcontentsline=\nocontentsline#1{#2}\egroup}

%%%%%%%%%%%%%%%%%%%%%%%%%%%%%%%%%%%%%%%%%%%%%%%%%%%%%%%%%%%%%%%%%%%%%%
\section{Near horizon expansion for Kerr} \label{Append:Kerr-expansion}
%%%%%%%%%%%%%%%%%%%%%%%%%%%%%%%%%%%%%%%%%%%%%%%%%%%%%%%%%%%%%%%%%%%%%%

Here we present the expansion of the Kerr geometry around an arbitrary point on the future horizon, i.e., we start from \eqref{eq:kerr1} and expand around $x^-=0$ for arbitrary (but finite) $x^+$. We display only non-zero metric components
\begin{align}
g_{++}&=\mathcal{O}\left([x^{-}]^{2}\right)\\
g_{+-}&= -2r_+(r_+ + r_-\cos^2\!\theta) - \frac{4r_+^2r_-\sin^2\!\theta}{r_+ + r_-\cos^2\!\theta}\,x^+x^-  +\mathcal{O}\left([x^{-}]^{2}\right) \\
g_{--}&=\frac{2r_+(r_+^2-2r_-(r_++r_-)\cos^2\!\theta-r_-^2\cos^4\theta)}{r_+ + r_-\cos^2\!\theta}\,x^{+\,2} + F(\theta)\, x^{+\,3}x^- +  \mathcal{O}\left([x^{-}]^{2}\right)\\
g_{+\phi}&= -\sin^2\!\theta\,H(\theta)\,x^- + \mathcal{O}\left([x^{-}]^{2}\right)\\
g_{-\phi}&= \sin^2\!\theta\,H(\theta)\,x^+ - (r_+-r_-)\,\sin^2\!\theta\,K(\theta)\,x^{+\,2}x^- +\mathcal{O}\left([x^{-}]^{2}\right)\\
g_{\phi \phi}&= \frac{r_+(r_++r_-)^2\sin^2\!\theta}{r_+ + r_-\cos^2\!\theta} - (r_+-r_-)\sin^2\!\theta\,G(\theta)\,x^+x^- + \mathcal{O}\left([x^{-}]^{2}\right) \\
g_{\theta \theta}&= r_+(r_+ + r_- \cos^2\!\theta) - 2 r_+ (r_+ - r_-)x^+x^- +\mathcal{O}\left([x^{-}]^{2}\right)
\end{align}
with the definitions 
\begin{align}
F(\theta) &= r_+^2-2r_-^2-r_+r_-\cos^2\!\theta - 8r_+r_-\cos^2\!\theta\,\frac{r_+^2-r_-^2}{(r_+ + r_-\cos^2\!\theta)^2}\\
G(\theta) &= r_+-r_- + \frac{r_+(r_+^2-r_-^2)+r_-(3r_+^2+4r_+r_-+r_-^2)\cos^2\!\theta}{(r_+ + r_-\cos^2\!\theta)^2} \\
H(\theta) &= \sqrt{r_+r_-}\,(r_+ - r_-) + 2\sqrt{r_+r_-}\,\frac{r_+\,\left(r_+ + r_-\right)}{r_++r_-\cos^2\!\theta} \\
K(\theta) &= \sqrt{r_+r_-} - (r_+ + r_-)\,\sqrt{r_+r_-}\,\frac{r_+ - 3r_-\cos^2\!\theta}{(r_++r_-\cos^2\!\theta)^2}\,.
\end{align}

%For Schwarzschild black holes ($r_-=0$) the expansion above simplifies considerably (again only non-zero metric coefficients are displayed):
%\begin{align} 
%g_{++}&=\mathcal{O}\left([x^{-}]^{2}\right)\\
%g_{+-}&= -2r_+^2 +\mathcal{O}\left([x^{-}]^{2}\right) \\
%g_{--}&= 2r_+^2\,x^{+\,2} + r_+^2\, x^{+\,3}x^- +  \mathcal{O}\left([x^{-}]^{2}\right)\\
%g_{\phi \phi}&= r_+^2\sin^2\!\theta - 2r_+^2\sin^2\!\theta\,x^+x^- + \mathcal{O}\left([x^{-}]^{2}\right) \\
%%g_{+\phi}&= g_{-\phi}= g_{\theta\phi}0\\
%g_{\theta \theta}&= r_+^2 - 2 r_+^2 x^+x^- +\mathcal{O}\left([x^{-}]^{2}\right)
%\end{align}

Comparing the results above with the fall-off conditions \eqref{NFH01} yields 
\begin{subequations}
\label{Kerr-pert}
\begin{align}
{\cal G} &= {\cal U}_\theta = {\cal V}_\theta = \Omega_{\theta\phi} = 0 & \frac{{\cal F}}{x^{+\,2}} &= \frac{4r_+^2(r_+ + r_-)}{r_+ + r_-\cos^2\!\theta} -2r_+\big(r_+ + r_-(2+\cos^2\!\theta)\big)\\
\eta &= -2r_+(r_+ + r_-\cos^2\!\theta) & \frac{{\cal U}_\phi}{x^+} &= - {\cal V}_\phi\ = \sin^2\!\theta\, H(\theta)\\
\Omega_{\theta\theta} &=r_+(r_+ + r_-\cos^2\!\theta) & \Omega_{\phi\phi} &= \frac{r_+(r_++r_-)^2\sin^2\!\theta}{r_+ + r_-\cos^2\!\theta}\,.
\end{align}
\end{subequations}

%%%%%%%%%%%%%%%%%%%%%%%%%%%%%%%%%%%%%%%%%%%%%%%%%%%%%%%%%%%%%%%%%%%%%%
\section{Some subalgebras of the {\NHS} algebra}\label{Appen:B}  
%%%%%%%%%%%%%%%%%%%%%%%%%%%%%%%%%%%%%%%%%%%%%%%%%%%%%%%%%%%%%%%%%%%%%%

The {\NHS} algebra \eqref{eq:angelinajolie} has different subalgebras which on their own may arise as near horizon symmetry algebras for more restrictive sets of boundary conditions (like in section \ref{sec:7}). In this appendix we list some of these subalgebras. 

\tocless\subsection{Finite subalgebras} 

Finite subalgebras of the near horizon/asymptotic symmetry algebras are typically associated with the global symmetries (isometries) of the corresponding background or their conformal extensions. Finite maximal subalgebras are subalgebras where no additional generator can be added without producing an infinite algebra. 

Rather than being encyclopedic we focus on the physically most interesting finite maximal subalgebra, where the superrotations are restricted to ${\cal L}^a$, ${\cal \bar{L}}^a$, $a=\pm1, 0$, which form an sl$(2,\mathbb{R})\oplus\,$sl$(2,\mathbb{R})\simeq\,$so$(2,2)$ algebra. Geometrically, this algebra is the global part of the conformal algebra acting on the bifurcation surface ${\cal B}$. The supertranslations are restricted to ${\cal R}^{0,0}_a, \ a=\pm1, 0$ and ${\cal T}^{0,0}_0$. These generators commute with the conformal algebra and form an $sl(2,\mathbb{R})\oplus u(1)$ subalgebra. Geometrically, ${\cal R}_{-1}^{0,0}$ generates rigid timeshifts, $x^+\to x^+ + x^+_0$, while ${\cal R}^{0,0}_0$ and ${\cal T}^{0,0}_0$ generate rigid scalings (also known as conformal boosts) in $x^+$ and $x^-$ directions, respectively, i.e., $x^+\to \lambda_+ x^+$ and $x^-\to \lambda_- x^-$. Therefore, ${\cal R}^{0,0}_0 -{\cal T}^{0,0}_0$ generates boosts and ${\cal R}^{0,0}_0 + {\cal T}^{0,0}_0$ dilatations on the $x^\pm$ plane. The remaining generator ${\cal R}_1^{0,0}$ generates special conformal translations along $x^+$. Notably, we do not have translations (nor special conformal translations) along $x^-$ and hence the 2d Poincar\'e algebra is not contained in our subalgebra. Physically, this is expected since we fix the locus of the null hypersurface to be $x^-=0$, a condition which manifestly breaks translation invariance along $x^-$.

In conclusion, an interesting finite maximal subalgebra of the {\NHS} algebra \eqref{eq:angelinajolie} is so$(2,2)\oplus\,$sl$(2,\mathbb{R})\oplus\,$u$(1)$, with the first part corresponding to global conformal transformations of the bifurcation surface, the second part to a chiral half of global conformal transformations in the $x^\pm$ plane and the abelian factor corresponding to boosts along $x^-$. As differential operators the generators of this algebra are given by $\partial_z$, $z\partial_z$, $z^2\partial_z$, $\partial_{\bar z}$, $\bar z\partial_{\bar z}$, $\bar z^2\partial_{\bar z}$, $\partial_+$, $x^+\partial_+$, $(x^+)^2\partial_+$ and $x^-\partial_-$.

\tocless\subsection{Infinite subalgebras} 

One can identify many different infinite dimensional subalgebras of the {\NHS} algebra \eqref{eq:angelinajolie}. Here we list some of the most important ones.
\begin{enumerate}
    \item One can switch off everything except for the T-Witt algebra generated by ${\cal R}_n^{p,q}$ and ${\cal T}_n^{p,q}$, keeping rigid the bifurcation surface $\cal B$. The T-Witt algebra itself also has some interesting subalgebras, for instance the BMS$_3^{(s=0)}$ algebra (also known as `warped Witt algebra') generated by ${\cal R}_n:={\cal R}_n^{0,0}$ as Witt generators and ${\cal T}_n:={\cal T}_n^{0,0}$ as lower-spin supertranslations, or the sl$(2,\mathbb{R})\oplus u(1)$ extended Witt algebra generated by  ${\cal R}_{a}^{p,q}, a=\pm1, 0$ and ${\cal T}_{0}^{p,q}$.
    \item Alternatively, one can switch off the T-Witt part of the algebra. The remainder is a Witt$\,\oplus\,$Witt subalgebra, the conformal algebra over ${\cal B}$, generated by ${\cal L}^p$, ${\cal{\bar L}}^p$.
    \item Physically more relevant are combined subalgebras that contain part of the T-Witt algebra and the full conformal algebra over ${\cal B}$. For example, one can keep the T-Witt generators ${\cal R}_n^{0,0}$, ${\cal T}_n^{0,0}$ and all Witt generators ${\cal L}^p$, ${\cal {\bar L}}^q$, yielding the subalgebra Witt$\,\oplus\,$Witt$\,\oplus$\,BMS$_3^{(s=0)}$. This subalgebra is an infinite-dimensional lift of the maximal finite subalgebra discussed above. In the remaining cases below we single out some of these algebras that either appeared in earlier literature or in the present work by imposing additional restrictions on the boundary conditions. We start with the former.
    \item There are infinitely many DGGP type \cite{Donnay:2015abr} subalgebras, generated by ${\cal L}^p$, ${\cal {\bar L}}^p$, ${\cal R}_n^{p,q}$ or ${\cal L}^p$, ${\cal {\bar L}}^p$, ${\cal T}_n^{p,q}$ for any fixed $n$. These algebras are BMS$_4^{(s=0)}$ in the notation of \cite{Grumiller:2019fmp}, or ${\cal W}(0,0;0,0)$ in the notation of \cite{Safari:2019zmc}. The only difference to the original BMS$_4$ algebra \cite{Bondi:1962,Sachs:1962,Barnich:2011ct} is the lower spin of the supertranslations, see the discussion in the appendix of \cite{Grumiller:2019fmp}.
    \item The algebra discussed in \cite{Donnay:2016ejv, Chandrasekaran:2018aop, Chandrasekaran:2019ewn} is a subalgebra generated by ${\cal L}^p$, ${\cal {\bar L}}^p$, ${\cal R}_{-1}^{p,q}$ and ${\cal R}_0^{p,q}$. Upon changing the basis\footnote{%
    More generally, the algebra automorphism ${\cal R}_n^{p,q} \to {\cal R}_n^{p,q} + \alpha (n+1) {\cal T}_n^{p,q}$ (with some constant $\alpha$) amounts to shifting the {\NHS} generator $T^- \to  T^- + \alpha \partial_+ T^+$.
    } ${\cal R}_0^{p,q}\to{\cal R}_0^{p,q}+{\cal T}_0^{p,q}$ this subalgebra arises as a special case discussed in section \ref{sec:7} where surface gravity is constant and Gaussian null coordinates are used (with loss of generality), see the next item. 
    \item Finally, we address subalgebras that emerged as special cases in sections \ref{sec:onshellphasespace} or \ref{sec:7}. The non-expanding case discussed in section \ref{sec:6.3} yields a subalgebra generated by ${\cal L}^p$, $\bar{\cal L}^p$, ${\cal T}^{p,q}_0$ and ${\cal R}_n^{p,q}$, see the statements after \eqref{Field-Variations-Stationary}. Constant surface gravity with \eqref{kappa-l-constant-GNC} leads to a subalgebra generated by ${\cal L}^p$, $\bar{\cal L}^p$, ${\cal T}^{p,q}_n$, ${\cal R}_{-1}^{p,q}$ and ${\cal R}_0^{p,q}$. Gaussian null coordinates, i.e., constant $\eta$, lead to a subalgebra generated by ${\cal L}^p$, $\bar{\cal L}^p$ and ${\cal R}^{p,q}_n$, see the statements below \eqref{Tpm-sec-6.2.2'}. Combining both cases, i.e., assuming constant $\eta$ and constant surface gravity, leads to a smaller subalgebra, generated by ${\cal L}^p$, $\bar{\cal L}^p$, ${\cal R}_{-1}^{p,q}$ and ${\cal R}_0^{p,q}+{\cal T}_0^{p,q}$, see the statements below \eqref{DGGP-Vector'}. Killing horizons in general also yield the constant surface gravity subalgebra, but when restricted to near bifurcation boundary conditions recover the smaller algebra \eqref{NHA-B} found in section \ref{sec:3}, which is a subalgebra generated by ${\cal L}^p$, $\bar{\cal L}^p$, ${\cal T}_0^{p,q}$ and ${\cal R}_0^{p,q}$. [Notational alert: in \eqref{NHA-B} all mode indices appear as lower indices and are labelled by letters $n,m,l,k$, as opposed to \eqref{eq:angelinajolie}.] Since only the linear combination ${\cal R}_0^{p,q}+{\cal T}_0^{p,q}$ enters the charges this subalgebra further reduces to the DGGP algebra \cite{Donnay:2015abr}, generated by ${\cal L}^p$, $\bar{\cal L}^p$ and supertranslations ${\cal R}_0^{p,q}+{\cal T}_0^{p,q}$.
\end{enumerate}

%\addcontentsline{toc}{section}{References}
%\bibliographystyle{fullsort.bst}
%\bibliography{reference}

\begin{thebibliography}{10}

\bibitem{Kerr:1963ud}
R.~P. Kerr, ``Gravitational field of a spinning mass as an example of
  algebraically special metrics,'' {\em Phys. Rev. Lett.} {\bf 11} (1963)
237--238.
%%CITATION = PRLTA,11,237;%%.

\bibitem{Einstein:1915ca}
A.~Einstein, ``Die {F}eldgleichungen der {G}ravitation,'' {\em Sitzungsber.
  Preuss. Akad. Wiss. Berlin (Math. Phys.)} {\bf 1915} (1915)
844--847.
%%CITATION = SPWPA,1915,844;%%.

\bibitem{Donnay:2015abr}
L.~Donnay, G.~Giribet, H.~A. Gonz{\'a}lez, and M.~Pino, ``{Supertranslations
  and Superrotations at the Black Hole Horizon},'' {\em Phys. Rev. Lett.} {\bf
  116} (2016), no.~9, 091101,
\href{http://www.arXiv.org/abs/1511.08687}{{\tt 1511.08687}}.
%%CITATION = ARXIV:1511.08687;%%.

\bibitem{Donnay:2016ejv}
L.~Donnay, G.~Giribet, H.~A. Gonz{\'a}lez, and M.~Pino, ``{Extended Symmetries
  at the Black Hole Horizon},'' {\em JHEP} {\bf 09} (2016) 100,
\href{http://www.arXiv.org/abs/1607.05703}{{\tt 1607.05703}}.
%%CITATION = ARXIV:1607.05703;%%.

\bibitem{Chandrasekaran:2018aop}
V.~Chandrasekaran, {\'E}.~{\'E}. Flanagan, and K.~Prabhu, ``{Symmetries and
  charges of general relativity at null boundaries},'' {\em JHEP} {\bf 11}
  (2018) 125,
\href{http://www.arXiv.org/abs/1807.11499}{{\tt 1807.11499}}.
%%CITATION = ARXIV:1807.11499;%%.

\bibitem{Haco:2018ske}
S.~Haco, S.~W. Hawking, M.~J. Perry, and A.~Strominger, ``{Black Hole Entropy
  and Soft Hair},'' {\em JHEP} {\bf 12} (2018) 098,
\href{http://www.arXiv.org/abs/1810.01847}{{\tt 1810.01847}}.
%%CITATION = ARXIV:1810.01847;%%.

\bibitem{Grumiller:2019fmp}
D.~Grumiller, A.~Pérez, M.~M. Sheikh-Jabbari, R.~Troncoso, and C.~Zwikel,
  ``{Spacetime structure near generic horizons and soft hair},'' {\em Phys.
  Rev. Lett.} {\bf 124} (2020) 041601,
\href{http://www.arXiv.org/abs/1908.09833}{{\tt 1908.09833}}.
%%CITATION = ARXIV:1908.09833;%%.

\bibitem{Kruskal:1959vx}
M.~D. Kruskal, ``{Maximal extension of Schwarzschild metric},'' {\em Phys.
  Rev.} {\bf 119} (1960)
1743--1745.
%%CITATION = PHRVA,119,1743;%%.

\bibitem{Israel:1966}
W.~Israel, ``New interpretation of the extended {S}chwarzschild manifold,''
  {\em Phys.Rev.} {\bf 143} (1966) 1016.

\bibitem{Bardeen:1999px}
J.~M. Bardeen and G.~T. Horowitz, ``{The extreme Kerr throat geometry: A vacuum
  analog of AdS(2) x S(2)},'' {\em Phys. Rev.} {\bf D60} (1999) 104030,
\href{http://www.arXiv.org/abs/hep-th/9905099}{{\tt hep-th/9905099}}.
%%CITATION = HEP-TH/9905099;%%.

\bibitem{Barnich:2009se}
G.~Barnich and C.~Troessaert, ``{Symmetries of asymptotically flat 4
  dimensional spacetimes at null infinity revisited},'' {\em Phys. Rev. Lett.}
  {\bf 105} (2010) 111103,
\href{http://www.arXiv.org/abs/0909.2617}{{\tt 0909.2617}}.
%%CITATION = ARXIV:0909.2617;%%.

\bibitem{Barnich:2010eb}
G.~Barnich and C.~Troessaert, ``{Aspects of the BMS/CFT correspondence},'' {\em
  JHEP} {\bf 1005} (2010) 062,
\href{http://www.arXiv.org/abs/1001.1541}{{\tt 1001.1541}}.
%%CITATION = ARXIV:1001.1541;%%.

\bibitem{Regge:1974zd}
T.~Regge and C.~Teitelboim, ``Role of surface integrals in the {H}amiltonian
  formulation of general relativity,'' {\em Ann. Phys.} {\bf 88} (1974)
286.
%%CITATION = APNYA,88,286;%%.

\bibitem{Brown:1986nw}
J.~D. Brown and M.~Henneaux, ``{Central Charges in the Canonical Realization of
  Asymptotic Symmetries: An Example from Three-Dimensional Gravity},'' {\em
  Commun. Math. Phys.} {\bf 104} (1986)
207--226.
%%CITATION = CMPHA,104,207;%%.

\bibitem{Henneaux:2018cst}
M.~Henneaux and C.~Troessaert, ``{BMS Group at Spatial Infinity: the
  Hamiltonian (ADM) approach},'' {\em JHEP} {\bf 03} (2018) 147,
\href{http://www.arXiv.org/abs/1801.03718}{{\tt 1801.03718}}.
%%CITATION = ARXIV:1801.03718;%%.

\bibitem{Lee:1990nz}
J.~Lee and R.~M. Wald, ``{Local symmetries and constraints},'' {\em J. Math.
  Phys.} {\bf 31} (1990)
725--743.
%%CITATION = JMAPA,31,725;%%.

\bibitem{Iyer:1994ys}
V.~Iyer and R.~M. Wald, ``Some properties of {N}{\"o}ther charge and a proposal
  for dynamical black hole entropy,'' {\em Phys. Rev.} {\bf D50} (1994)
  846--864,
\href{http://arXiv.org/abs/gr-qc/9403028}{{\tt gr-qc/9403028}}.
%%CITATION = GR-QC 9403028;%%.

\bibitem{Barnich:2001jy}
G.~Barnich and F.~Brandt, ``{Covariant theory of asymptotic symmetries,
  conservation laws and central charges},'' {\em Nucl. Phys.} {\bf B633} (2002)
  3--82,
\href{http://www.arXiv.org/abs/hep-th/0111246}{{\tt hep-th/0111246}}.
%%CITATION = HEP-TH/0111246;%%.

\bibitem{Barnich:2006av}
G.~Barnich and G.~Compere, ``{Classical central extension for asymptotic
  symmetries at null infinity in three spacetime dimensions},'' {\em
  Class.Quant.Grav.} {\bf 24} (2007) F15--F23,
\href{http://www.arXiv.org/abs/gr-qc/0610130}{{\tt gr-qc/0610130}}.
%%CITATION = GR-QC/0610130;%%.

\bibitem{Compere:2018aar}
G.~Compère and A.~Fiorucci, ``{Advanced Lectures on General Relativity},''
  {\em Lect. Notes Phys.} {\bf 952} (2019) 150,
\href{http://www.arXiv.org/abs/1801.07064}{{\tt 1801.07064}}.
%%CITATION = ARXIV:1801.07064;%%.

\bibitem{Ashtekar:1996cd}
A.~Ashtekar, J.~Bicak, and B.~G. Schmidt, ``{Asymptotic structure of symmetry
  reduced general relativity},'' {\em Phys.Rev.} {\bf D55} (1997) 669--686,
\href{http://www.arXiv.org/abs/gr-qc/9608042}{{\tt gr-qc/9608042}}.
%%CITATION = GR-QC/9608042;%%.

\bibitem{Parsa:2018kys}
A.~Farahmand~Parsa, H.~R. Safari, and M.~M. Sheikh-Jabbari, ``{On Rigidity of
  3d Asymptotic Symmetry Algebras},'' {\em JHEP} {\bf 03} (2019) 143,
\href{http://www.arXiv.org/abs/1809.08209}{{\tt 1809.08209}}.
%%CITATION = ARXIV:1809.08209;%%.

\bibitem{Barnich:2011ct}
G.~Barnich and C.~Troessaert, ``{Supertranslations call for superrotations},''
  {\em PoS} {\bf CNCFG2010} (2010) 010,
  \href{http://www.arXiv.org/abs/1102.4632}{{\tt 1102.4632}}.
[Ann. U. Craiova Phys.21,S11(2011)].
%%CITATION = ARXIV:1102.4632;%%.

\bibitem{Barnich:2011mi}
G.~Barnich and C.~Troessaert, ``{BMS charge algebra},'' {\em JHEP} {\bf 1112}
  (2011) 105,
\href{http://www.arXiv.org/abs/1106.0213}{{\tt 1106.0213}}.
%%CITATION = ARXIV:1106.0213;%%.

\bibitem{Compere:2015knw}
G.~Comp{\`e}re, P.-J. Mao, A.~Seraj, and M.~M. Sheikh-Jabbari, ``{Symplectic
  and Killing symmetries of AdS$_{3}$ gravity: holographic vs boundary
  gravitons},'' {\em JHEP} {\bf 01} (2016) 080,
\href{http://www.arXiv.org/abs/1511.06079}{{\tt 1511.06079}}.
%%CITATION = ARXIV:1511.06079;%%.

\bibitem{Bunster:2018yjr}
  C.~Bunster, A.~Gomberoff and A.~P{\'e}rez,
  ``Regge--Teitelboim analysis of the symmetries of electromagnetic and gravitational fields on asymptotically null spacelike surfaces,''
\href{http://www.arXiv.org/abs/1805.03728}{{\tt 1805.03728}}.

\bibitem{Chandrasekaran:2019ewn}
V.~Chandrasekaran and K.~Prabhu, ``{Symmetries, charges and conservation laws
  at causal diamonds in general relativity},'' {\em JHEP} {\bf 10} (2019) 229,
\href{http://www.arXiv.org/abs/1908.00017}{{\tt 1908.00017}}.
%%CITATION = ARXIV:1908.00017;%%.

\bibitem{Campiglia:2020qvc}
M.~Campiglia and J.~Peraza, ``{Generalized BMS charge algebra},''
\href{http://www.arXiv.org/abs/2002.06691}{{\tt 2002.06691}}.
%%CITATION = ARXIV:2002.06691;%%.

\bibitem{Strominger:2017zoo}
A.~Strominger, ``{Lectures on the Infrared Structure of Gravity and Gauge
  Theory},''
\href{http://www.arXiv.org/abs/1703.05448}{{\tt 1703.05448}}.
%%CITATION = ARXIV:1703.05448;%%.

\bibitem{Harlow:2019yfa}
D.~Harlow and J.-Q. Wu, ``{Covariant phase space with boundaries},''
\href{http://www.arXiv.org/abs/1906.08616}{{\tt 1906.08616}}.
%%CITATION = ARXIV:1906.08616;%%.

\bibitem{Geiller:2019bti}
M.~Geiller and P.~Jai-akson, ``{Extended actions, dynamics of edge modes, and
  entanglement entropy},''
\href{http://www.arXiv.org/abs/1912.06025}{{\tt 1912.06025}}.
%%CITATION = ARXIV:1912.06025;%%.

\bibitem{Barnich:2019xhd}
G.~Barnich, ``{Black hole entropy from nonproper gauge degrees of freedom: The
  charged vacuum capacitor},'' {\em Phys. Rev.} {\bf D99} (2019), no.~2,
  026007,
\href{http://www.arXiv.org/abs/1806.00549}{{\tt 1806.00549}}.
%%CITATION = ARXIV:1806.00549;%%.

\bibitem{Wald:1999wa}
R.~M. Wald and A.~Zoupas, ``{A General definition of 'conserved quantities' in
  general relativity and other theories of gravity},'' {\em Phys.Rev.} {\bf
  D61} (2000) 084027,
\href{http://www.arXiv.org/abs/gr-qc/9911095}{{\tt gr-qc/9911095}}.
%%CITATION = GR-QC/9911095;%%.

\bibitem{tHooft:1990fkf}
G.~'t~Hooft, ``{The black hole interpretation of string theory},'' {\em Nucl.
  Phys.} {\bf B335} (1990)
138--154.
%%CITATION = NUPHA,B335,138;%%.

\bibitem{tHooft:1991uqr}
G.~'t~Hooft, ``{The Black hole horizon as a quantum surface},'' {\em Phys.
  Scripta} {\bf T36} (1991)
247--252.
%%CITATION = PHSTB,T36,247;%%.

\bibitem{Susskind:1993if}
L.~Susskind, L.~Thorlacius, and J.~Uglum, ``{The Stretched horizon and black
  hole complementarity},'' {\em Phys.Rev.} {\bf D48} (1993) 3743--3761,
\href{http://www.arXiv.org/abs/hep-th/9306069}{{\tt hep-th/9306069}}.
%%CITATION = HEP-TH/9306069;%%.

\bibitem{Hayward:1993wb}
S.~A. Hayward, ``General laws of black hole dynamics,'' {\em Phys. Rev.} {\bf
  D49} (1994)
6467--6474.
%%CITATION = PHRVA,D49,6467;%%.

\bibitem{Carlip:1994gy}
S.~Carlip, ``The statistical mechanics of the (2+1)-dimensional black hole,''
  {\em Phys. Rev.} {\bf D51} (1995) 632--637,
\href{http://www.arXiv.org/abs/gr-qc/9409052}{{\tt gr-qc/9409052}}.
%%CITATION = GR-QC 9409052;%%.

\bibitem{Balachandran:1994up}
A.~P. Balachandran, L.~Chandar, and A.~Momen, ``Edge states in gravity and
  black hole physics,'' {\em Nucl. Phys.} {\bf B461} (1996) 581--596,
\href{http://www.arXiv.org/abs/gr-qc/9412019}{{\tt gr-qc/9412019}}.
%%CITATION = GR-QC 9412019;%%.

\bibitem{Carlip:1995cd}
S.~Carlip, ``{Statistical mechanics and black hole entropy},'' in {\em Field
  Theory, Integrable Systems and Symmetries}.
\newblock 1997.
\newblock
\href{http://www.arXiv.org/abs/gr-qc/9509024}{{\tt gr-qc/9509024}}.
\newblock
%%CITATION = GR-QC/9509024;%%.

\bibitem{Strominger:1997eq}
A.~Strominger, ``Black hole entropy from near-horizon microstates,'' {\em JHEP}
  {\bf 02} (1998) 009,
\href{http://www.arXiv.org/abs/hep-th/9712251}{{\tt hep-th/9712251}}.
%%CITATION = HEP-TH 9712251;%%.

\bibitem{Ashtekar:1997yu}
A.~Ashtekar, J.~Baez, A.~Corichi, and K.~Krasnov, ``Quantum geometry and black
  hole entropy,'' {\em Phys. Rev. Lett.} {\bf 80} (1998) 904--907,
\href{http://www.arXiv.org/abs/gr-qc/9710007}{{\tt gr-qc/9710007}}.
%%CITATION = GR-QC 9710007;%%.

\bibitem{Carlip:1998wz}
S.~Carlip, ``Black hole entropy from conformal field theory in any dimension,''
  {\em Phys. Rev. Lett.} {\bf 82} (1999) 2828--2831,
\href{http://www.arXiv.org/abs/hep-th/9812013}{{\tt hep-th/9812013}}.
%%CITATION = HEP-TH 9812013;%%.

\bibitem{Hotta:2000gx}
M.~Hotta, K.~Sasaki, and T.~Sasaki, ``{Diffeomorphism on horizon as an
  asymptotic isometry of Schwarzschild black hole},'' {\em Class. Quant. Grav.}
  {\bf 18} (2001) 1823--1834,
\href{http://www.arXiv.org/abs/gr-qc/0011043}{{\tt gr-qc/0011043}}.
%%CITATION = GR-QC/0011043;%%.

\bibitem{Ashtekar:2000sz}
A.~Ashtekar {\em et al.}, ``Isolated horizons and their applications,'' {\em
  Phys. Rev. Lett.} {\bf 85} (2000) 3564--3567,
\href{http://www.arXiv.org/abs/gr-qc/0006006}{{\tt gr-qc/0006006}}.
%%CITATION = GR-QC 0006006;%%.

\bibitem{Ashtekar:2002ag}
A.~Ashtekar and B.~Krishnan, ``{Dynamical horizons: Energy, angular momentum,
  fluxes and balance laws},'' {\em Phys. Rev. Lett.} {\bf 89} (2002) 261101,
\href{http://www.arXiv.org/abs/gr-qc/0207080}{{\tt gr-qc/0207080}}.
%%CITATION = GR-QC/0207080;%%.

\bibitem{tHooft:2006xjp}
G.~'t~Hooft, ``{The Black hole horizon as a dynamical system},'' {\em Int. J.
  Mod. Phys.} {\bf D15} (2006) 1587--1602,
\href{http://www.arXiv.org/abs/gr-qc/0606026}{{\tt gr-qc/0606026}}.
%%CITATION = GR-QC/0606026;%%.

\bibitem{Majhi:2012tf}
B.~R. Majhi and T.~Padmanabhan, ``{Noether current from the surface term of
  gravitational action, Virasoro algebra and horizon entropy},'' {\em Phys.
  Rev.} {\bf D86} (2012) 101501,
\href{http://www.arXiv.org/abs/1204.1422}{{\tt 1204.1422}}.
%%CITATION = ARXIV:1204.1422;%%.

\bibitem{Penna:2015gza}
R.~F. Penna, ``{BMS invariance and the membrane paradigm},'' {\em JHEP} {\bf
  03} (2016) 023,
\href{http://www.arXiv.org/abs/1508.06577}{{\tt 1508.06577}}.
%%CITATION = ARXIV:1508.06577;%%.

\bibitem{Afshar:2015wjm}
H.~Afshar, S.~Detournay, D.~Grumiller, and B.~Oblak, ``{Near-Horizon Geometry
  and Warped Conformal Symmetry},'' {\em JHEP} {\bf 03} (2016) 187,
\href{http://www.arXiv.org/abs/1512.08233}{{\tt 1512.08233}}.
%%CITATION = ARXIV:1512.08233;%%.

\bibitem{Hawking:2016msc}
S.~W. Hawking, M.~J. Perry, and A.~Strominger, ``{Soft Hair on Black Holes},''
  {\em Phys. Rev. Lett.} {\bf 116} (2016), no.~23, 231301,
\href{http://www.arXiv.org/abs/1601.00921}{{\tt 1601.00921}}.
%%CITATION = ARXIV:1601.00921;%%.

\bibitem{Hooft:2016itl}
G.~'t~Hooft, ``{Black hole unitarity and antipodal entanglement},'' {\em Found.
  Phys.} {\bf 46} (2016), no.~9, 1185--1198,
\href{http://www.arXiv.org/abs/1601.03447}{{\tt 1601.03447}}.
%%CITATION = ARXIV:1601.03447;%%.

\bibitem{Averin:2016ybl}
A.~Averin, G.~Dvali, C.~Gomez, and D.~Lust, ``{Gravitational Black Hole Hair
  from Event Horizon Supertranslations},'' {\em JHEP} {\bf 06} (2016) 088,
\href{http://www.arXiv.org/abs/1601.03725}{{\tt 1601.03725}}.
%%CITATION = ARXIV:1601.03725;%%.

\bibitem{Compere:2016hzt}
G.~Comp{\`e}re and J.~Long, ``{Classical static final state of collapse with
  supertranslation memory},'' {\em Class. Quant. Grav.} {\bf 33} (2016),
  no.~19, 195001,
\href{http://www.arXiv.org/abs/1602.05197}{{\tt 1602.05197}}.
%%CITATION = ARXIV:1602.05197;%%.

\bibitem{Afshar:2016wfy}
H.~Afshar, S.~Detournay, D.~Grumiller, W.~Merbis, A.~Perez, D.~Tempo, and
  R.~Troncoso, ``{Soft Heisenberg hair on black holes in three dimensions},''
  {\em Phys. Rev.} {\bf D93} (2016), no.~10, 101503,
\href{http://www.arXiv.org/abs/1603.04824}{{\tt 1603.04824}}.
%%CITATION = ARXIV:1603.04824;%%.

\bibitem{Eling:2016xlx}
C.~Eling and Y.~Oz, ``{On the Membrane Paradigm and Spontaneous Breaking of
  Horizon BMS Symmetries},'' {\em JHEP} {\bf 07} (2016) 065,
\href{http://www.arXiv.org/abs/1605.00183}{{\tt 1605.00183}}.
%%CITATION = ARXIV:1605.00183;%%.

\bibitem{Afshar:2016uax}
H.~Afshar, D.~Grumiller, and M.~M. Sheikh-Jabbari, ``{Near horizon soft hair as
  microstates of three dimensional black holes},'' {\em Phys. Rev.} {\bf D96}
  (2017), no.~8, 084032,
\href{http://www.arXiv.org/abs/1607.00009}{{\tt 1607.00009}}.
%%CITATION = ARXIV:1607.00009;%%.

\bibitem{Mirbabayi:2016axw}
M.~Mirbabayi and M.~Porrati, ``{Dressed Hard States and Black Hole Soft
  Hair},'' {\em Phys. Rev. Lett.} {\bf 117} (2016), no.~21, 211301,
\href{http://www.arXiv.org/abs/1607.03120}{{\tt 1607.03120}}.
%%CITATION = ARXIV:1607.03120;%%.

\bibitem{Hopfmuller:2016scf}
F.~Hopfmüller and L.~Freidel, ``{Gravity Degrees of Freedom on a Null
  Surface},'' {\em Phys. Rev.} {\bf D95} (2017), no.~10, 104006,
\href{http://www.arXiv.org/abs/1611.03096}{{\tt 1611.03096}}.
%%CITATION = ARXIV:1611.03096;%%.

\bibitem{Hawking:2016sgy}
S.~W. Hawking, M.~J. Perry, and A.~Strominger, ``{Superrotation Charge and
  Supertranslation Hair on Black Holes},'' {\em JHEP} {\bf 05} (2017) 161,
\href{http://www.arXiv.org/abs/1611.09175}{{\tt 1611.09175}}.
%%CITATION = ARXIV:1611.09175;%%.

\bibitem{Afshar:2016kjj}
H.~Afshar, D.~Grumiller, W.~Merbis, A.~Perez, D.~Tempo, and R.~Troncoso,
  ``{Soft hairy horizons in three spacetime dimensions},'' {\em Phys. Rev.}
  {\bf D95} (2017), no.~10, 106005,
\href{http://www.arXiv.org/abs/1611.09783}{{\tt 1611.09783}}.
%%CITATION = ARXIV:1611.09783;%%.

\bibitem{Wieland:2017zkf}
W.~Wieland, ``{New boundary variables for classical and quantum gravity on a
  null surface},'' {\em Class. Quant. Grav.} {\bf 34} (2017), no.~21, 215008,
\href{http://www.arXiv.org/abs/1704.07391}{{\tt 1704.07391}}.
%%CITATION = ARXIV:1704.07391;%%.

\bibitem{Bousso:2017dny}
R.~Bousso and M.~Porrati, ``{Soft Hair as a Soft Wig},'' {\em Class. Quant.
  Grav.} {\bf 34} (2017), no.~20, 204001,
\href{http://www.arXiv.org/abs/1706.00436}{{\tt 1706.00436}}.
%%CITATION = ARXIV:1706.00436;%%.

\bibitem{Strominger:2017aeh}
A.~Strominger, ``{Black Hole Information Revisited},'' pp.~109--117.
\newblock 2020.
\newblock
\href{http://www.arXiv.org/abs/1706.07143}{{\tt 1706.07143}}.
\newblock
%%CITATION = ARXIV:1706.07143;%%.

\bibitem{Akhmedov:2017ftb}
E.~T. Akhmedov and M.~Godazgar, ``{Symmetries at the black hole horizon},''
  {\em Phys. Rev.} {\bf D96} (2017), no.~10, 104025,
\href{http://www.arXiv.org/abs/1707.05517}{{\tt 1707.05517}}.
%%CITATION = ARXIV:1707.05517;%%.

\bibitem{Lust:2017gez}
D.~Lust, ``{Supertranslations and Holography near the Horizon of Schwarzschild
  Black Holes},'' {\em Fortsch. Phys.} {\bf 66} (2018), no.~2, 1800001,
\href{http://www.arXiv.org/abs/1711.04582}{{\tt 1711.04582}}.
%%CITATION = ARXIV:1711.04582;%%.

\bibitem{Blommaert:2018rsf}
A.~Blommaert, T.~G. Mertens, H.~Verschelde, and V.~I. Zakharov, ``{Edge State
  Quantization: Vector Fields in Rindler},'' {\em JHEP} {\bf 08} (2018) 196,
\href{http://www.arXiv.org/abs/1801.09910}{{\tt 1801.09910}}.
%%CITATION = ARXIV:1801.09910;%%.

\bibitem{Hooft:2018gtw}
G.~'t~Hooft, ``{What happens in a black hole when a particle meets its
  antipode},''
\href{http://www.arXiv.org/abs/1804.05744}{{\tt 1804.05744}}.
%%CITATION = ARXIV:1804.05744;%%.

\bibitem{Donnelly:2018nbv}
W.~Donnelly and S.~B. Giddings, ``{Gravitational splitting at first order:
  Quantum information localization in gravity},'' {\em Phys. Rev.} {\bf D98}
  (2018), no.~8, 086006,
\href{http://www.arXiv.org/abs/1805.11095}{{\tt 1805.11095}}.
%%CITATION = ARXIV:1805.11095;%%.

\bibitem{Donnay:2018ckb}
L.~Donnay, G.~Giribet, H.~A. Gonz{\'a}lez, and A.~Puhm, ``{Black hole memory
  effect},'' {\em Phys. Rev.} {\bf D98} (2018), no.~12, 124016,
\href{http://www.arXiv.org/abs/1809.07266}{{\tt 1809.07266}}.
%%CITATION = ARXIV:1809.07266;%%.

\bibitem{Penna:2018gfx}
R.~F. Penna, ``{Near-horizon Carroll symmetry and black hole Love numbers},''
\href{http://www.arXiv.org/abs/1812.05643}{{\tt 1812.05643}}.
%%CITATION = ARXIV:1812.05643;%%.

\bibitem{Donnay:2019jiz}
L.~Donnay and C.~Marteau, ``{Carrollian Physics at the Black Hole Horizon},''
\href{http://www.arXiv.org/abs/1903.09654}{{\tt 1903.09654}}.
%%CITATION = ARXIV:1903.09654;%%.

\bibitem{Wieland:2019hkz}
W.~Wieland, ``{Generating functional for gravitational null initial data},''
  {\em Class. Quant. Grav.} {\bf 36} (2019), no.~23, 235007,
\href{http://www.arXiv.org/abs/1905.06357}{{\tt 1905.06357}}.
%%CITATION = ARXIV:1905.06357;%%.

\bibitem{Grumiller:2019tyl}
D.~Grumiller and W.~Merbis, ``{Near horizon dynamics of three dimensional black
  holes},'' {\em SciPost Phys.} {\bf 8} (2020) 010,
\href{http://www.arXiv.org/abs/1906.10694}{{\tt 1906.10694}}.
%%CITATION = ARXIV:1906.10694;%%.

\bibitem{Grumiller:2019ygj}
D.~Grumiller, M.~M. Sheikh-Jabbari, C.~Troessaert, and R.~Wutte,
  ``{Interpolating Between Asymptotic and Near Horizon Symmetries},''
\href{http://www.arXiv.org/abs/1911.04503}{{\tt 1911.04503}}.
%%CITATION = ARXIV:1911.04503;%%.

\bibitem{Bagchi:2019clu}
A.~Bagchi, R.~Basu, A.~Mehra, and P.~Nandi, ``{Field Theories on Null
  Manifolds},''
\href{http://www.arXiv.org/abs/1912.09388}{{\tt 1912.09388}}.
%%CITATION = ARXIV:1912.09388;%%.

\bibitem{Ashtekar:2020ifw}
A.~Ashtekar, ``{Black Hole evaporation: A Perspective from Loop Quantum
  Gravity},'' {\em Universe} {\bf 6} (2020), no.~2, 21,
\href{http://www.arXiv.org/abs/2001.08833}{{\tt 2001.08833}}.
%%CITATION = ARXIV:2001.08833;%%.

\bibitem{Hajian:2015xlp}
K.~Hajian and M.~M. Sheikh-Jabbari, ``{Solution Phase Space and Conserved
  Charges: A General Formulation for Charges Associated with Exact
  Symmetries},'' {\em Phys. Rev.} {\bf D93} (2016), no.~4, 044074,
\href{http://www.arXiv.org/abs/1512.05584}{{\tt 1512.05584}}.
%%CITATION = ARXIV:1512.05584;%%.

\bibitem{Guica:2008mu}
M.~Guica, T.~Hartman, W.~Song, and A.~Strominger, ``{The Kerr/CFT
  Correspondence},'' {\em Phys. Rev.} {\bf D80} (2009) 124008,
\href{http://www.arXiv.org/abs/0809.4266}{{\tt 0809.4266}}.
%%CITATION = ARXIV:0809.4266;%%.

\bibitem{Compere:2012jk}
G.~Compere, ``{The Kerr/CFT correspondence and its extensions: a comprehensive
  review},'' {\em Living Rev.Rel.} {\bf 15} (2012) 11,
\href{http://www.arXiv.org/abs/1203.3561}{{\tt 1203.3561}}.
%%CITATION = ARXIV:1203.3561;%%.

\bibitem{Aggarwal:2019iay}
  A.~Aggarwal, A.~Castro and S.~Detournay,
  ``Warped Symmetries of the Kerr Black Hole,''
  {\em JHEP} {\bf 01} (2020) 016, \href{http://www.arXiv.org/abs/1909.03137}{{\tt 1909.03137}}.

\bibitem{Flanagan:2019vbl}
E.~E. Flanagan, K.~Prabhu, and I.~Shehzad, ``{Extensions of the asymptotic
  symmetry algebra of general relativity},'' {\em JHEP} {\bf 01} (2020) 002,
\href{http://www.arXiv.org/abs/1910.04557}{{\tt 1910.04557}}.
%%CITATION = ARXIV:1910.04557;%%.

\bibitem{Afshar:2017okz}
H.~Afshar, D.~Grumiller, M.~M. Sheikh-Jabbari, and H.~Yavartanoo, ``{Horizon
  fluff, semi-classical black hole microstates --- Log-corrections to BTZ
  entropy and black hole/particle correspondence},'' {\em JHEP} {\bf 08} (2017)
  087,
\href{http://www.arXiv.org/abs/1705.06257}{{\tt 1705.06257}}.
%%CITATION = ARXIV:1705.06257;%%.

\enlargethispage{1truecm}

\bibitem{Grumiller:2016pqb}
D.~Grumiller and M.~Riegler, ``{Most general AdS$_{3}$ boundary conditions},''
  {\em JHEP} {\bf 10} (2016) 023,
\href{http://www.arXiv.org/abs/1608.01308}{{\tt 1608.01308}}.
%%CITATION = ARXIV:1608.01308;%%.

\bibitem{Grumiller:2017sjh}
D.~Grumiller, W.~Merbis, and M.~Riegler, ``{Most general flat space boundary
  conditions in three-dimensional Einstein gravity},'' {\em Class. Quant.
  Grav.} {\bf 34} (2017), no.~18, 184001,
\href{http://www.arXiv.org/abs/1704.07419}{{\tt 1704.07419}}.
%%CITATION = ARXIV:1704.07419;%%.

\bibitem{Strominger:2013jfa}
A.~Strominger, ``{On BMS Invariance of Gravitational Scattering},'' {\em JHEP}
  {\bf 07} (2014) 152,
\href{http://www.arXiv.org/abs/1312.2229}{{\tt 1312.2229}}.
%%CITATION = ARXIV:1312.2229;%%.

\bibitem{Pasterski:2019msg}
S.~Pasterski, ``{Implications of Superrotations},'' {\em Phys. Rept.} {\bf 829}
  (2019) 1--35,
\href{http://www.arXiv.org/abs/1905.10052}{{\tt 1905.10052}}.
%%CITATION = ARXIV:1905.10052;%%.

\bibitem{Safari:2019zmc}
H.~R. Safari and M.~M. Sheikh-Jabbari, ``{BMS$_{4}$ algebra, its stability and
  deformations},'' {\em JHEP} {\bf 04} (2019) 068,
\href{http://www.arXiv.org/abs/1902.03260}{{\tt 1902.03260}}.
%%CITATION = ARXIV:1902.03260;%%.

\bibitem{Bondi:1962}
H.~Bondi, M.~van~der Burg, and A.~Metzner, ``Gravitational waves in general
  relativity {VII.} {W}aves from axi-symmetric isolated systems,'' {\em Proc.
  Roy. Soc. London} {\bf A269} (1962) 21--51.

\bibitem{Sachs:1962}
R.~Sachs, ``Asymptotic symmetries in gravitational theory,'' {\em Phys. Rev.}
  {\bf 128} (1962) 2851--2864.

\end{thebibliography}

\providecommand{\href}[2]{#2}\begingroup\raggedright\endgroup

\end{document}